\documentclass[11pt,onecolumn]{IEEEtran}

\pdfoutput=1
\usepackage{graphicx}
\usepackage{amssymb}
\usepackage{amsmath}
\usepackage{amsthm}
\usepackage{dsfont}
\usepackage{url}
\usepackage{tikz}
\newtheorem{theorem}{Theorem}

\newtheorem*{lemnonum}{Lemma}
\newtheorem{prop}[theorem]{Proposition}
\newtheorem{defn}{Definition}

\newcommand{\norm}[1]{\left\Vert#1\right\Vert}
\newcommand{\abs}[1]{\left\vert#1\right\vert}
\newcommand{\set}[1]{\left\{#1\right\}}

\newcommand{\ceil}[1]{\left\lceil #1 \right\rceil }

\newcommand{\Reals}{\mathds R}
\newcommand{\Integers}{\mathds Z}

\newcommand{\Prob}[1]{\mathrm{Pr}\left\{#1\right\}}

\newcommand{\bx}{\mathbf{x}}

\newcommand{\bb}{\mathbf{b}}

\newcommand{\by}{\mathbf{y}}
\newcommand{\bz}{\mathbf{z}}
\newcommand{\bs}{\mathbf{s}}
\newcommand{\bv}{\mathbf{v}}

\newcommand{\bX}{\mathbf{X}}
\newcommand{\bY}{\mathbf{Y}}

\newcommand{\bS}{\mathbf{S}}
\newcommand{\bE}{\mathbf{E}}
\newcommand{\bM}{\mathbf{M}}

\newcommand{\cC}{\mathcal{C}}

\newcommand{\cB}{\mathcal{B}}

\newcommand{\cS}{\mathcal{S}}

\newcommand{\bu}{\mathbf{u}}

\newcommand{\ra}{\rightarrow}

\newcommand{\id}[1]{\,\mathrm{d}#1}

\usepackage{tikz}
\usepackage{xcolor}
\usepackage{loops}
\usetikzlibrary{plotmarks}

\usepackage{color}
\usepackage{import}

\newcommand{\ID}{\mathrm{ID}}
\newcommand{\CAP}{\mathrm{CAP}}

\newcommand{\CONE}{\mathrm{CONE}}

\newcommand{\COV}{\mathrm{COV}}

\newcommand{\MAYBE}{\mathtt{maybe}}

\newcommand{\NO}{\mathtt{no}}

\usepackage{pgfplots}
\pgfplotsset{compat=newest}

\usepackage{capt-of}
\usepackage{mathtools}
\usepackage{dsfont}
\usepackage[ruled,vlined]{algorithm2e}

\usetikzlibrary{calc,positioning,shapes.geometric}
\newcommand*\circled[1]{\tikz[baseline=(char.base)]{\node[shape=circle,draw,inner sep=2pt] (char) {#1};}}

\graphicspath{{figures/}}

\usepackage{pgfplots}
\usepackage{tikz}
\usepackage{capt-of}
\usetikzlibrary{arrows,intersections}
\usepackage{rotating}
\pgfplotsset{compat=newest}

\newdimen\mydim
 \newcommand\getx[1]{
      \pgfextractx\mydim{\pgfpointanchor{#1}{center}}
 }
 \newdimen\mydimm
 \newcommand\getxx[1]{
      \pgfextractx\mydimm{\pgfpointanchor{#1}{center}}
 }
  \newdimen\mydimmm
 \newcommand\getxxx[1]{
      \pgfextractx\mydimmm{\pgfpointanchor{#1}{center}}
 }

\definecolor{qqwuqq}{rgb}{0.13,0.13,0.13}
\definecolor{qqqqff}{rgb}{0.33,0.33,0.33}
\definecolor{xdxdff}{rgb}{0.66,0.66,0.66}
\definecolor{uuuuuu}{rgb}{0.27,0.27,0.27}

\begin{document}

\title{Compression for Quadratic Similarity Queries: Finite Blocklength and Practical Schemes}
\author{Fabian Steiner, Steffen Dempfle, Amir Ingber and Tsachy Weissman

\thanks{Fabian Steiner and Steffen Dempfle are with the Dept. of Electrical Engineering of the Technische Universit\"at M\"unchen, 
Munich, Germany. Email: \{fabian.steiner,steffen.dempfle\}@tum.de}
\thanks{Amir Ingber was with the Dept. of Electrical Engineering, Stanford University, Stanford, CA 94305. He is now with Yahoo! Labs, Sunnyvale, CA 94089 Email: ingber@yahoo-inc.com.}
\thanks{Tsachy Weissman is with the Dept. of Electrical Engineering, Stanford University, Stanford, CA 94305. 
Email: tsachy@stanford.edu.}
\thanks{Parts of this work will be presented at the 2014 IEEE International Symposium on Information Theory (ISIT).}
}

\maketitle

\begin{abstract}

We study the problem of compression for the purpose of similarity identification, where similarity is measured 
by the mean square Euclidean distance between vectors. While the asymptotical fundamental limits of the problem -- the minimal 
compression rate and the error exponent -- were found in a previous work, in this paper we focus on the nonasymptotic domain
and on practical, implementable schemes.

We first present a finite blocklength achievability bound based on shape-gain quantization: The gain (amplitude) of the vector 
is compressed via scalar quantization and the shape (the projection on the unit sphere) is quantized using a spherical code. 
The results are numerically evaluated and they converge to the asymptotic values as predicted by the error exponent.
We then give a nonasymptotic lower bound on the performance of any compression scheme, and compare to the upper (achievability) 
bound. For a practical implementation of such a scheme, we use wrapped spherical codes, studied by Hamkins and Zeger, and use 
the Leech lattice as an example for an underlying lattice. As a side result, we obtain a bound on the covering angle of 
any wrapped spherical code, as a function of the covering radius of the underlying lattice.
\end{abstract}

\section{Introduction}
\label{sec:intro}

%
%

The number of applications dealing with a huge amount of data has increased significantly in recent years. Many of
those applications do not only deal with storage of the data, but also with its retrieval and querying.
In many cases, the query is whether a given database contains sequences that are
similar to a given sequence of interest. The notion of ``similarity'' depends on the kind of application involved, where 
notable examples include the Hamming and Euclidean distances. The size of the big database motivates the question of how to 
construct a much smaller (compressed) version of it that will allow to answer queries reliably.

In \cite{ingberSQCD} Ingber et al. develop a general framework for the case of a Gaussian source and Euclidean distance 
measure and they also provide asymptotic results for the identification rate, i.e. the rate above which any query can be made 
arbitrarily reliable, and a characterization of the identification exponent associated with it. Results for the case
of discrete memoryless sources are given in \cite{IW13_IT}.

In the present work, we follow the framework described in \cite{ingberSQCD} and extend it to the 
finite blocklength case. We begin by deriving a nonasymptotic achievability bound on the reliability, using shape-gain 
quantizers \cite{gersho1992}. In such systems, the gain (amplitude) of the vector is compressed via scalar quantization and 
the shape (the projection on the unit sphere) is quantized using a spherical code. While in \cite{ingberSQCD} the asymptotics of the setting 
allow crude scalar quantizers, here we optimize the quantizers for the distribution of the source. Combined with a 
(nonconstructive) result on the covering of spherical shells \cite{Dumer07}, the performance of the system can be evaluated 
numerically at any finite blocklength $n$. The numerical result validates the asymptotic approximations for the performance 
predicted by the error exponent of \cite{ingberSQCD}. The achievability result is complemented by a lower bound on 
the performance at finite blocklength. The lower bound is derived following the approach in \cite{ingberSQCD}, but with 
greater attention to detail and optimization of the different parameters involved in the derivation.

In addition to the (non-constructive) achievability result, we develop a general method of 
constructing implementable compression schemes, which are also based on the shape-gain framework. 
While the gain quantizer of the achievability bound can be easily implemented, the shape quantizer (the spherical code) 
is not. For that purpose, we utilize \emph{wrapped} spherical codes which were previously introduced in 
\cite{hamkinsWrappedCodes}. The shape codebook is obtained by considering a mapping which wraps an $n-1$-dimensional 
lattice around the shell of the $n$-dimensional unit sphere. Any lattice can be used for this process and its covering 
radius defines the performance of the scheme. As part of the analysis of the scheme, we derive a bound on the covering 
angle of any wrapped spherical code (as a function of the properties of the underlying lattice), a result that may be of 
independent interest.

The rest of this paper is organized as follows. The next subsections introduce terms and definitions
that are used throughout the paper. Section~\ref{sec:achievability} presents the achievabilty results, whereas
Section~\ref{sec:lb} is dedicated to the converse result. Section~\ref{sec:practical_implementation}
describes an actual, implementable scheme that can be used in practice, along with numerical results. We will provide some
concluding remarks and possible further research objectives in Section~\ref{sec:remarks}.

\subsection{Problem Setting}
\label{sec:problem_setting}

The goal of the framework presented in \cite{ingberSQCD} is to answer similarity queries from a compressed 
representation of the data. 
More specifically, for each sequence $\bx$ in the database, we only keep a compressed
signature $Q(\bx)$. The final goal is to be able to detect whether $\bx$ is similar to a query sequence $\by$,
given only $Q(\bx)$ and $\by$.

\begin{center}
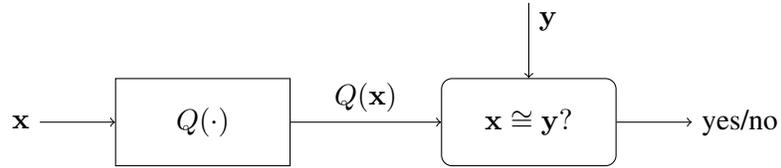

\begin{tikzpicture} 
 \tikzstyle{block} = [draw, rectangle, minimum height=3em, minimum width=6em];
 \node (x){$\bx$};
 \node[block,right= 1cm of x] (sig) {$Q(\cdot)$};
 \node[block,rounded corners,right= 2cm of sig] (dec) {$\bx \cong \by?$};
 \draw[->] (x) -- (sig);
 \draw[->] (sig) -- (dec) node[above,midway] {$Q(\bx)$};
 \draw[->] (dec.east) -- +(1,0) node[right] {yes/no};
 \draw[->] (dec.north) +(0,1) -- (dec.north) node[right,near start] {$\by$};
\end{tikzpicture}
\captionof{figure}{Answering a query from compressed data.}
\label{fig:query}
\end{center}

Concerning the nature of the answer ``yes/no'' of the setup depicted in Figure~\ref{fig:query}, the possible errors are either 
false positives or false negatives. While the first event is not considered catastrophic, as it only results in additional efforts when the answer of the original query has to be confirmed with the actual database entry 
in addition to its compressed version, the incident of false negatives can not be detected: Many practical applications, e.g. 
querying a criminal forensic database, obviously need to exclude this kind of error.

Therefore, we impose the restriction to our model that \emph{false negatives are not permitted}. Basically, this means that the 
result of the query function is either ``$\NO$" or ``$\MAYBE$", where the latter pertains to the cases of being either 
actually similar or false positive.

We focus on a similarity measure defined by the normalized squared Euclidean distance. To this end, for any length-$n$ real 
sequences $\bx = \left(x_1,x_2,\ldots,x_n\right)^T, \by=\left(y_1,y_2,\ldots,y_n\right)^T \in \Reals^n$ define

\begin{align}
  d(\bx,\by) \triangleq \frac{1}{n}\sum_{i=1}^n (x_i-y_i)^2 = \frac{1}{n}\|\bx-\by\|^2,\label{eq:def_distance}
\end{align}
where $\|\cdot\|$ denotes the standard Euclidean norm. We say that $\bx$ and $\by$ are $D$-\emph{similar} when 
$d(\bx,\by)\leq D$, or simply \emph{similar} when $D$ is clear from the context.

To formalize the previously described problem setting, we define the following (see \cite{ingberSQCD}): 

\begin{defn}
A rate-$R$ identification system $(Q,g)$ consists of a \emph{signature assignment}
\begin{align}
Q : \Reals^n \ra \{1,2,\dots,2^{nR}\}
\end{align}
and a \emph{query function}
\begin{align}
g : \{1,2,\dots,2^{nR}\}\times \Reals^n \ra \{\NO, \MAYBE\}.
\end{align}
\end{defn}

\begin{defn}\label{def:d_admissible}
A system $(Q,g)$ is said to be $D$-\emph{admissible}, if for any $\bx,\by$ satisfying $d(\bx,\by)\leq D$, we have
\begin{equation}\label{eqn:maybe}
  g(Q(\bx),\by) = \MAYBE.
\end{equation}
\end{defn}

Note that, by definition, any $D$-{admissible} system $(Q,g)$ can not produce false negatives.

At this point, it is worthwhile to think about the figure of merit that should be considered in our
system design. In the spirit of source and channel coding scenarios, where one generally aims at driving the
error probability to zero for long blocklengths, we pursue the same idea
with the probability of a false positive event $\mathcal{E} = \{ g(Q(\bX),\bY) = \MAYBE,  d(\bX,\bY) > D\}$. Assuming a
$D$-admissible system $(Q,g)$ we can relate this probability to $\Pr \{ g(Q(\bX),\bY) = \MAYBE \}$ and concern ourselves with the latter
quantity instead:

\begin{align}
\Prob{g(Q(\bX),\bY) = \MAYBE}
&= \Prob{ g(Q(\bX),\bY) = \MAYBE | d(\bX,\bY)\leq D\}\Pr\{ d(\bX,\bY)\leq D} \notag\\
&\quad+ \Prob{g(Q(\bX),\bY) = \MAYBE,  d(\bX,\bY) > D} \notag\\
&=\Prob{d(\bX,\bY)\leq D} + \Prob{\mathcal{E}}, \label{eq:maybeRelation}
\end{align}

\noindent where \eqref{eq:maybeRelation} follows since $\Pr \{ g(Q(\bX),\bY) = \MAYBE | d(\bX,\bY)\leq D\}=1$ by the $D$-admissibility 
of $(Q,g)$.  Since $\Pr\{ d(\bX,\bY)\leq D\}$ does not depend on what scheme is employed, minimizing the false positive 
probability $\Pr\{\mathcal{E}\}$ over all $D$-admissible schemes $(Q,g)$ is equivalent to minimizing 
$\Pr \{ g(Q(\bX),\bY) = \MAYBE \}$. Also note that the only interesting case is 
when $\Pr\{d(\bX,\bY)\leq D \} \ra 0$ as $n$ grows, since otherwise almost all the sequences in the 
database will be similar to the query sequence, making the problem degenerate (since almost all the 
database needs to be retrieved, regardless of the compression). In this case, it is easy to see 
that $\Pr\{\mathcal{E}\}$ vanishes if and only if the conditional probability

\begin{equation}
  \Pr \{ g(Q(\bX),\bY) = \MAYBE | d(\bX,\bY)> D\}
\end{equation}
vanishes. In view of the above, we henceforth restrict our attention to the behavior of $\Pr \{ g(Q(\bX),\bY) = \MAYBE \}$.  

In analogy to the classical rate-distortion setting \cite{GallagerIT,CoverThomasIT}, we also define:

\begin{defn}
    For given distributions $P_X, P_Y$ and a similarity threshold $D$, a rate $R$ is said to be $D$-\emph{achievable} if there exists a sequence of  rate-$R$ admissible schemes 
    $(Q^{(n)},g^{(n)})$ satisfying
    \begin{equation}
      \lim_{n\ra\infty} \Pr\left\{g^{(n)}\left(Q^{(n)}(\bX),\bY \right) = \MAYBE\right\} = 0, \label{eqn:reliable}
    \end{equation}
\end{defn}
where $\bX$ and $\bY$ are independent i.i.d. sequences with respective marginals $P_X$ and $P_Y$.

\begin{defn}
    For given distributions $P_X, P_Y$ and a similarity threshold $D$, the \emph{identification rate}  $R_\ID(D,P_X,P_Y)$ is the infimum of $D$-achievable rates.  That is,
   \begin{align}
     R_\ID(D,P_X,P_Y) \triangleq \inf \{ R : R~ \mbox{is $D$-achievable}\}\label{eq:def_id_rate},
   \end{align}
   where an infimum over the empty set is equal to $\infty$.
\end{defn}

One can also define the identification exponent, i.e. the asymptotic slope of the exponential decay of $\Pr \{ g(Q(\bX),\bY) = \MAYBE \}$:

\begin{defn}
    Fix  $R\geq R_\ID(D,P_X,P_Y)$. The \emph{identification exponent} is defined as 
    \begin{align}
        \bE_\ID(R,D,P_X,P_Y) \triangleq \limsup_{n \ra \infty} -\frac{1}{n}\log \inf_{g^{(n)},Q^{(n)}} \Pr \left\{ g^{(n)}\left(Q^{(n)}(\bX),\bY \right) = \MAYBE \right\},\label{eq:def_exp}
    \end{align}
    where the infimum is over all $D$-admissible systems $(g^{(n)},Q^{(n)})$ of rate $R$ and blocklength $n$.
\end{defn}

Note that this quantity gives rise to the approximation $\Prob{\MAYBE} \approx e^{-n\bE_\ID(R)}$, assuming an approximately optimal scheme is employed, which is valid for large $n$.

In the following, we will focus on the standard Gaussian case, meaning that the components $X_1,\ldots,X_n$ and $Y_1,\ldots,Y_n$
of the length-$n$ vectors $\bX$ and $\bY$ are independent and identically distributed Gaussian random variables with zero mean and
unit variance. For this special (but important) case, the identification rate is given by \cite[Corollary 1]{ingberSQCD}

\begin{align}
      R_\ID(D) = \left\{
      \begin{array}{ll}
      \log\left(\frac{2}{2-D}\right)\quad \quad &\mbox{for $0 \leq D < 2$}\\
      \infty & \mbox{for $D \geq 2$}.
      \end{array}
      \right.
       \label{eq:R_ID_gauss}
\end{align}

The exponent for this case is given by \cite[Corollary 2]{ingberSQCD}

\begin{align}
	\mathbf{E}_{\text{ID}}(R,D) & = \min_{\rho} \frac{1}{\ln 2} (\rho-1 -\ln{\rho}) - \log{\sin{\min{\left[\frac{\pi}{2}, \left(\arcsin{(2^{-R})} + \arccos{\frac{2\rho - D}{2\rho}} \right) \right]}}}\\
				    & \text{s.t } 2 \geq 2\rho\geq D\nonumber.
\end{align}

\subsection{Geometry Basics Revisited}
\label{sec:spherical_geometry}

For the derivations in the next sections some results from Euclidean geometry are needed. The reason for 
this is mainly due to the fact that we concentrate on Gaussian random vectors and the distribution of the shape $\bX/\norm{\bX}$ of a Gaussian vector $\bX$ is uniform on the shell of the unit sphere in $n$ dimensions. 
More generally, we define the spherical shell with arbitrary radius $r > 0$ as
\begin{align}
 \cS^n_r \triangleq \set{\bx\in\Reals^n: \norm{\bx} = r}.\label{eq:def_sphere}
\end{align}

If the index $r$ is omitted for notational brevity, we shall refer to the spherical unit shell. In case the interior should
be part of the set as well, we speak of a ball that is usually centered around a point $\bu\in\Reals^n$:
\begin{align}
 \cB_r(\bu) \triangleq \set{\bx\in\Reals^n: \norm{\bu-\bx} \leq r}.
\end{align}

The definition of a spherical shell can be extended to a ``thick'' spherical shell in $n$ dimensions by
\begin{align}
 \cS^n_{r_1,r_2}\triangleq \set{\bx\in\Reals^n: r_1 \leq \norm{\bx} \leq r_2}.\label{eq:def_shell}
\end{align}

The angle between two elements $\bx_1$, $\bx_2$ can be expressed as:
\begin{align}
 \angle(\bx_1,\bx_2) \triangleq \arccos\left(\frac{\bx_1^T\bx_2}{\norm{\bx_1}\norm{\bx_2}}\right) \in [0,\pi].\label{eq:def_angle}
\end{align}

Given a point $\bu\in\Reals^n\backslash\set{0}$ and half angle $\theta\in[0,\pi]$, define:
\begin{align}
 \CONE(\bu,\theta) \triangleq \set{\bx\in\Reals^n: \angle(\bx,\bu) \leq \theta}.\label{eq:def_cone}
\end{align}

The definitions of~\eqref{eq:def_sphere} and \eqref{eq:def_cone} now become vital as the intersection of the two describes a
spherical cap denoted by $\CAP_r(\bu,\theta)$:
\begin{align}
 \CAP_r(\bu,\theta) \triangleq \cS^n_r \cap \CONE(\bu,\theta)\label{eq:def_cap}.
\end{align}

Employing the notion of a thick shell in~\eqref{eq:def_shell}, we can also define a thick cap given as:
\begin{align}
 \CAP_{r_1,r_2}(\bu,\theta) \triangleq \cS^n_{r_1,r_2} \cap \CONE(\bu,\theta)\label{eq:def_thick_cap}.
\end{align}

When talking about coverings of spherical shells as needed for quantization purposes, we will need to compute their $n-1$ and $n$-dimensional
contents. According to \cite{li2011}, these are calculated as
\begin{align}
 \abs{\cS_r^n} &= \frac{2\pi^{\frac{n}{2}}r^{n-1}}{\Gamma\left(\frac{n}{2}\right)},\label{eq:def_area_sphere}\\
 V(\cS_r^n) &= \frac{\pi^{\frac{n}{2}}r^n}{\Gamma\left(\frac{n+2}{2}\right)}\label{eq:def_vol_sphere}.
\end{align}

Note that the fraction of a spherical shell $\cS_r^n$ that is covered by a spherical cap $\CAP_{r}(\bu,\theta)$ 
can be expressed as
\begin{equation}
 \Omega(\theta,n) \triangleq \frac{|\CAP_r(\bu,\theta)|}{|\cS^n_r|} = \frac{1}{2} I_{\sin^2(\theta)}\left(\frac{n-1}{2}, \frac{1}{2}\right)\label{eq:def_omega},
\end{equation}

where the function $I_x(a,b) $ denotes the regularized, incomplete beta function:
\begin{equation}
 I_x(a,b) \triangleq \frac{\int_0^x t^a (1-t)^b \id{t}}{\int_0^1 t^a (1-t)^b \id{t}}.
\end{equation}

We emphasize that equation~\eqref{eq:def_omega} is solely dependent on the angle $\theta$ and the dimension $n$, but not on
the point $\bu$ or the radius $r$. This fact will facilitate the calculation of quantities of interest in 
Section~\ref{sec:achievability}. If the dimension $n$ is clear from the context, we omit the second parameter and simply 
write $\Omega(\theta)$.

Finally, for $A \subseteq \Reals^n$ and $D > 0$, we define the $D$-expansion of $A$ as
\begin{equation}
\Gamma^D(A) \triangleq \set{\by\in\Reals^n: \exists_{\bx\in A} d(\bx,\by) \leq D},\label{eq:gamma_exten}
\end{equation}

where $d(\bx,\by)$ was defined in \eqref{eq:def_distance}.

\subsection{Coverings and Lattices}
\label{sec:sphere_packing_lattice}

In this subsection, we introduce the general notion of coverings of a set and then show how these definitons directly apply
to coverings of lattices and spherical shells.

\subsubsection{Coverings}

Let $A\subseteq\Reals^n$, then we say that a set $B$ $\rho$-covers the set $A$, if
\begin{align}
 A \subseteq \bigcup_{\bx\in B} \cB_\rho(\bx)\label{eq:def_covering}.
\end{align}

We denote the collection of all sets that $\rho$-cover the set $A$ by $\COV(A,\rho)$. A convenient measure which allows for 
comparison between different coverings $B$ is provided by their covering density:
\begin{align}
 \zeta(A,B) \triangleq \sum_{\bx\in B} \frac{\abs{A\cap \cB_{\rho}(\bx)}}{\abs{A}}.\label{eq:def_density}
\end{align}

A classical task is to look for a covering $B \in \COV(A,\rho)$ which results in the smallest density. Formally, it can be found 
when \eqref{eq:def_density} is minimized over all coverings in $\COV(A,\rho)$:

\begin{align}
 \vartheta(A) \triangleq \min_{B \in \COV(A,\rho)} \zeta(A,B)\label{eq:def_density_minimal}.
\end{align}

As we have to quantize the shape of a Gaussian vector, which lies on the shell of the unit sphere, the set $A$ can be replaced by $\cS^n$
for our purposes. In this case, the intersection $\cS^n\cap \cB_\rho(\bx)$ results in a spherical cap,
namely $\CAP_1(\bu,\theta)$. Using this for the evaluation of \eqref{eq:def_density_minimal}, the quantity turns out to read as

\begin{align}
 \vartheta(\cS^n) = \abs{B^*}\cdot \Omega(\theta)\label{eq:def_density_spherical_code},
\end{align}
which states the covering density of a spherical code with covering angle $\theta$ and $B^*$ is the minimizer of
\eqref{eq:def_density_minimal}.

In \cite[Theorem 1]{Dumer07}, Dumer showed that $\vartheta(\cS^n)$ is bounded by
\begin{align}
 \vartheta(\cS^n) \leq (n-1)\log_2\left(n-1\right)\left(\frac{1}{2}+\frac{2\log_2\log_2\left(n-1\right)+5}{\log_2\left(n-1\right)}\right).\label{eq:bound_dumer}
\end{align}

We use this result in Section~\ref{sec:achievability} in order to retrieve an upper bound on the rate $R_S$ of the spherical code.


\subsubsection{Lattices}
\label{sec:lattices}

A lattice $\Lambda$ is a set of vectors  that is closed under addition, i.e. forms an additive group. It can be defined by 
a set of basis vectors  $\bv_1,\bv_2,\ldots,\bv_n \in \Reals^n$, i.e.

\begin{align}
 \Lambda \triangleq \set{\bv\in\Reals^n: \bv = \sum_{i=1}^n c_i \bv_i, c_i\in\Integers}\label{eq:def_lambda}
\end{align}

Generally, these vectors are combined in the generator matrix $\bM$ of the lattice $\Lambda$:
\begin{align}
 \bM = \begin{pmatrix}
        \bv_1 \quad\bv_2 \quad\ldots \quad\bv_n
       \end{pmatrix}.\label{eq:def_lambda_mtx}
\end{align}

Apart from the generator matrix, another important property is given by its minimum distance $d_\Lambda$
\begin{align}
 d_\Lambda = \min_{\substack{\bv,\bu\in\Lambda\\\bv\neq\bu}} \norm{\bv-\bu}\label{eq:def_lattice_min_d}
\end{align}

and covering radius $r_\Lambda^\mathrm{cov}$
\begin{align}
 r_{\Lambda}^\mathrm{cov} = \adjustlimits\sup_{\bx\in\Reals^n} \inf_{\bv\in\Lambda} \norm{\bx-\bv}\label{eq:def_lattice_cov_r}.
\end{align}

While the minimum distance $d_\Lambda$ is more important for channel coding applications (as it always ensures a certain distance between
two points in the lattice), the latter quantity is of special interest for source coding problems. It is reasonable to define the
packing radius as $r_\Lambda^\mathrm{pack} = d_\Lambda/2$, i.e. the largest radius balls that are centered
around any lattice point can have so that they do not overlap. Obviously, it always holds that $r_\Lambda^\mathrm{pack} \leq r_\Lambda^\mathrm{cov}$.

For a general lattice the last two quantities can not be easily determined, but require a deep geometrical understanding of the 
lattice. However, Sloane \cite{sloane_1981} has compiled a detailed comparison of the many important lattices used in practice,
and some approximative, numerical approaches can be found in the literature \cite{schuermann2006}, \cite[Chapter 2, 1.4]{ConwaySloane1993} as well.

Using $\bM$ as a description of a lattice has many advantages, as several important properties can be calculated easily.
One important operation that can be performed on a lattice involves the rescaling of its basis vectors $\bv_i, i \in [1:n]$.
A comprehensive survey of the effects of such a scaling is given in \cite{sloane_1981}. Assuming that $\Lambda' = s\Lambda, s \in\Reals$
is the scaled lattice, we particularly note that because of \eqref{eq:def_lattice_min_d} the minimum distance of the new lattice 
scales with the same factor $s$, i.e. $d_{\Lambda'} = sd_\Lambda$. We exploit this property in Section~\ref{sec:practical_implementation}
to adapt the rate of the wrapped spherical code quantizer.

\section{Achievability}
\label{sec:achievability}

%
%

\subsection{Proposed Achievability Scheme}
\label{sec:prop_scheme}

We have pointed out in Section~\ref{sec:problem_setting} that we wish to minimize $\Prob{g(Q(\bX)),\bY) = \MAYBE}$, or  
$\Prob{\MAYBE}$ for short, as it can be regarded as a performance measure of our scheme. In order for a scheme to be admissible 
according to Definition~\ref{def:d_admissible}, we must answer $\MAYBE$ whenever $\bY \in \Gamma^D(Q^{-1}(Q(\bX)))$, where 
we define the set of all the points that have a signature equal to $i$ as $Q^{-1}(i) \triangleq \set{\bx\in\Reals^n: Q(\bx) = i}$. 
Evidently, the corresponding probability can be written formally as

\begin{equation}
  \Prob{\MAYBE} = \sum_{i\in\left[1:2^{nR}\right]} \Prob{Q(\bX) = i} \Prob{\bY \in \Gamma^D\left(Q^{-1}\left(i\right)\right)}.\label{eq:prob_achiev1}
\end{equation}

Analyzing this quantity turns out to be a diffcult task, when no further structure or knowledge about the compression
scheme $Q(\cdot)$ is available. For that purpose, we shall construct  $Q(\cdot)$ as a shape-gain quantizer \cite{gersho1992}.

Shape-gain vector quantizers can be understood as a special implementation of product quantizers. The decomposition of the 
random vector $\bX$ is obtained by splitting it into its shape $\bS = \bX/\norm{\bX}$ and gain (amplitude) $G = \norm{\bX}$. 

For our case of a Gaussian random vector $\bX$, the random variable $G$ is a scalar value that follows a $\chi$-distribution 
\cite{miller1964} with $n$ degrees of freedom and possesses the probability density function

\begin{align}
 f_G(r) = f_{\norm{\bX}}(r) = \frac{2^{1-\frac{n}{2}r^{n-1}}\mathrm{e}^{-\frac{r^2}{2}}}{\Gamma\left(\frac{n}{2}\right)},\label{eq:def_chi}
\end{align}
where $\Gamma(n)$ denotes the usual gamma function: $\Gamma(n) = \int_{0}^\infty t^{(n-1)}\mathrm{e}^{-t}\id{t}$.
It is quantized via $Q_G: r \in \Reals^+_0 \to \left[1:2^{nR_G}\right]$, which can efficiently be realized by the Lloyd-Max 
algorithm \cite{llyod1982}, \cite{max1960}\footnote{While the Lloyd-Max is known to be optimal in some cases for MSE quantization, we have no optimality guarantee when used for compression for similarity 
identification. Nevertheless, as shown in the next sections, the performance is very good.}. We denote the boundaries of the 
quantization intervals as $\left[r_{k-1},r_k\right], k\in\left[1:2^{nR_G}\right]$.

We quantize the shape $\bS$ independently from the gain via a spherical code $Q_S: \cS^n \to \left[1:2^{nR_S}\right]$ and obtain the 
shape codebook $\cC_S$: As $\bS$ is now an element of the shell of the unit sphere, it is easier to quantize than the original 
random variable. In particular, because of the Gaussian assumption on $\bX$, the shape $\bS$ is uniformly distributed on the shell of the unit 
sphere (cf. \eqref{eq:uniform_on_sphere}), which motivates its quantization with a spherical code, which can be implemented, for 
example, by wrapping lattices in $\Reals^{n-1}$ around the spherical shell in $n$ dimensions \cite{hamkinsPhd}, a path we pursue in 
Section~\ref{sec:practical_implementation}.

\tikzstyle{block} = [draw, rectangle, minimum height=3em, minimum width=6em]
\tikzstyle{sum} = [draw, circle, node distance=1cm]
\tikzstyle{input} = [coordinate]
\tikzstyle{output} = [coordinate]
\tikzstyle{pinstyle} = [pin edge={to-,thin,black}]
\begin{center}
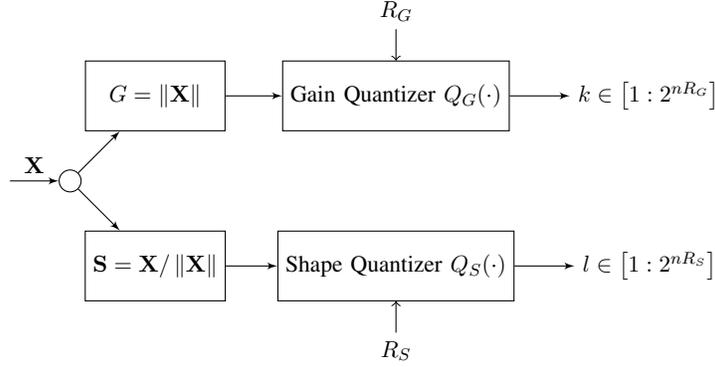

\begin{tikzpicture}[auto, node distance=2cm,>=latex',scale=.8,every node/.style={scale=0.8}]
    \node [input, name=input] {};
    \node [sum, right of=input] (split) {};
    \node [block, above right of=split] (G) {$G=\norm{\bX}$};
    \node [block, right of=G, pin={[pinstyle]above:$R_G$},
            node distance=4cm] (gain_q) {Gain Quantizer $Q_G(\cdot)$}; 
		\node [block, below right of=split] (S) {$\bS=\bX / \norm{\bX}$};
		\node [block, right of=S, pin={[pinstyle]below:$R_S$},
            node distance=4cm] (shape_q) {Shape Quantizer $Q_S(\cdot)$}; 
    \draw [draw,->] (input) -- node {$\bX$} (split);
    \draw [->] (split) -- (G);
		\draw [->] (split) -- (S);
		\draw [->] (G) -- (gain_q);
		\draw [->] (S) -- (shape_q);
    \draw[->] (gain_q.east) -- +(1,0) node[right] {$k\in\left[1:2^{nR_G}\right]$};
    \draw[->] (shape_q.east) -- +(1,0) node[right] {$l \in \left[1:2^{nR_S}\right]$};
\end{tikzpicture}
\captionof{figure}{Illustration of a shape-gain quantization process for the proposed achievability part.}
\label{fig:shape_gain_quant}
\end{center}

It is important to stress that we do not care about the exact reconstruction $\hat{\bX}=\hat{G}\hat{\bS}$ for which the
knowledge about the associated codebooks is essential, but are only interested in how close our query is to any point in the 
quantization cell. Nevertheless, we use the notation $\hat{\bs}$ to refer to  the center of a spherical cap that 
contains the quantization cell.

In order to prove the asymptotic achievability results, involving the identification rate, \cite{ingberSQCD} shows
that it is sufficient to neglect the gain quantizer and to concentrate on the ``typical gain'', as for high dimension $n$ the 
probability densitity function of $\bX$ concentrates near a hyperspherical shell $\cS^n_{r_X^-,r_X^+}$ with $r_X^\pm = \sqrt{n(\sigma_X^2\pm\eta)}$
and $\eta$ being arbitrarily small. In the nonasymptotic domain, we can no longer rely on this fact.

\subsection{Analysis of the Proposed Scheme}
\label{sec:analysis_scheme}

In \eqref{eq:prob_achiev1} we pointed out that the definition of our achievability scheme requires to answer $\MAYBE$ whenever
$\bY \in \Gamma^D(Q^{-1}(Q(\bX)))$. We can now find an upper bound for this probability by taking the structure of
a shape-gain quantizer into account. As before, we define the suggestive sets $Q_G^{-1}(k) \triangleq \set{r\in\Reals^+_0: Q_G(r) = k}= \left[r_{k-1},r_k\right]$
and $Q_S^{-1}(l) \triangleq \set{\bs\in\cS^n: Q_S(\bs) = l}$.

Note that it is trivial to embed both $k$ and $l$ in a single integer~$i$. Therefore, for a shape-gain quantizer, we have 
\begin{align}
 Q^{-1}(i) \triangleq \set{\left.\Reals^n\ni\bx = r\cdot\bs \right| r \in Q_G^{-1}(k), \bs \in Q_S^{-1}(l)}.\label{eq:quant_inv}
\end{align}

At this point, we only allow shape codebooks $\cC_S$ that come with a guaranteed upper bound on the covering angle 
$\angle(\bS,\hat{\bS}) \leq \theta$. Consequently, we conclude that the set $Q^{-1}(i)$ is contained within a thick cap,
i.e. $Q^{-1}(i) \subseteq \CAP_{r_{k-1},r_k}(\hat{\bs}_l,\theta)$. This fact simplifies the analysis and gives rise
to an easy implementation of an admissible decision rule, i.e. the second factor in \eqref{eq:prob_achiev1} can be written more 
explicitly by checking for $\by \in \Gamma^D\left(\CAP_{r_{k-1},r_k}(\hat{\bs_l},\theta)\right)$.

\begin{center}
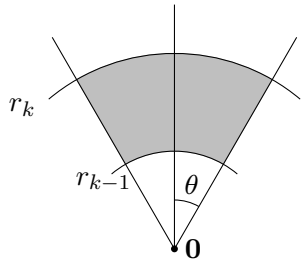

\begin{tikzpicture}[scale=1.3]


\fill (0,0) circle[radius=1pt] node[right] {$\mathbf{0}$};
\filldraw[gray!50] (60:1) -- (60:2) arc (60:120:2) -- (120:1) arc (120:60:1);
\draw (50:1) arc (50:130:1) node[xshift=-3pt,yshift=-3pt] {$r_{k-1}$};
\draw (50:2) arc (50:130:2) node[xshift=-10pt,yshift=-3pt] {$r_{k}$};

\draw (0,0) -- (90:2.5);
\draw (60:.5) arc (60:90:.5) node[midway,above,xshift=5pt,yshift=12pt] {\scriptsize $\theta$};
\draw (0,0) -- (60:2.5);
\draw (0,0) -- (120:2.5);

\end{tikzpicture}
\captionof{figure}{Illustration of the spherical cap that contains the set $Q^{-1}(i)$.}
\label{fig:set_qinv}
\end{center}

Hence, the upper bound on $\Prob{\MAYBE}$ is given by

 \begin{multline}
  \Prob{\MAYBE} \leq \sum_{k\in\left[1:2^{nR_G}\right]}\sum_{l\in\left[1:2^{nR_S}\right]} \Prob{Q_G(\norm{\bX})=k}\Prob{Q_S(\bS) = l}\Prob{\bY \in \Gamma^D\left(\CAP_{r_{k-1},r_k}(\hat{\bs}_l,\theta)\right)}\label{eq:prob_maybe_1}.
\end{multline}

The propositions that follow are geared toward obtaining simpler expressions for the 
last factor in $\eqref{eq:prob_maybe_1}$. The 
expression will turn out not to depend on a specific codepoint $\hat{\bs}_l$, so that we can also write 
with any $\hat{\bs} \in \cC_S$:

 \begin{equation}
  \Prob{\MAYBE} \leq \sum_{k\in\left[1:2^{nR_G}\right]} \Prob{\norm{\bX} \in\left[r_{k-1},r_{k}\right]} \Prob{\bY \in \Gamma^D\left(\CAP_{r_{k-1},r_k}(\hat{\bs},\theta)\right)}\label{eq:prob_maybe}.  
\end{equation}

Before calculating the probability of $\bY$ falling into the expansion of a thick cap as suggested by \eqref{eq:prob_maybe}, 
we approach this problem by first assuming that the gain quantization is a trivial mapping that maps the gain to one 
single value~$r$.

\vspace{2\baselineskip}
 \begin{prop}\label{prop:prop_thin_cap}
  The probability of the random variable $\bY$ falling into the $D$-expansion of a \emph{thin} spherical cap $\CAP_r(\hat{\bs},\theta)$ is given by
  \begin{equation}
   \Prob{\bY \in \Gamma^D(\CAP_r(\hat{\bs}, \theta)} = \int\limits_{0}^\infty \!\Prob{\bY \in \Gamma^D(\CAP_{r}(\hat{\bs}, \theta)) \left|\right. \norm{\bY} = r_Y} \cdot f_{\norm{\bY}}(r_Y)\id{r_Y}\label{eq:prob_y_thin_cap}
  \end{equation}
  
  and

  \begin{equation}
   \Prob{\bY \in \Gamma^D(\CAP_r(\hat{\bs}, \theta)) \left|\right. \norm{\bY} = r_Y} =\\ \begin{cases}
                                                                                             0, & |r-r_Y| > \sqrt{nD}\\
                                                                                             1, & r_Y \leq r_{Y, \text{deg}}(r,\theta) \\
                                                                                             \Omega(\theta+\theta'(r_Y)), & \text{otherwise}.\\
                                                                                          \end{cases}
  \end{equation}
  
  The quantity $r_{Y, \text{deg}}(r,\theta)$ is given by
  \begin{equation}
   r_{Y,\text{deg}}(r,\theta) =  \sqrt{(r\cos(\theta))^2-r^2+nD} - r\cos(\theta)
  \end{equation}
  and $\theta'(r_Y)$ by
  \begin{equation}
    \theta'(r_Y) = \arccos\left(\frac{r^2+r_Y^2-nD}{2\cdot r\cdot r_Y}\right).
  \end{equation} 
  
 \end{prop}
 
 \textit{Proof:} See Appendix~\ref{sec:proof_prop_1}.\hfill$\blacksquare$


 Using similar techniques, the analysis can be extended to a thick spherical cap as follows.
 
\begin{prop}\label{prop:prop_thick_cap}
  The probability of the random variable $\bY$ falling into the $D$-expansion of a \emph{thick} spherical cap $\CAP_{r_1,r_2}(\hat{\bs},\theta)$ is given by
  \begin{equation}
   \Prob{\bY \in \Gamma^D(\CAP_{r_1,r_2}(\hat{\bs}, \theta)} = \int\limits_{0}^\infty \!\Pr\left\{\bY \in \Gamma^D(\CAP_{r_1,r_2}(\hat{\bs}, \theta)) \left|\right. \norm{\bY} = r_Y\right\} \cdot f_{\norm{\bY}}(r_Y)\id{r_Y}
  \end{equation} and

  \begin{multline}
   \Prob{\bY \in \Gamma^D(\CAP_{r_1,r_2}(\hat{\bs}, \theta)) \left|\right. \norm{\bY} = r_Y} = \begin{cases}
											      0, & r_Y < r_1 - \sqrt{nD} \text{ or}\\
											         & r_Y > r_2 + \sqrt{nD} \\
                                                                                             1, & r_Y \leq r_{Y, \text{deg}}'(r_1, \theta) \\
                                                                                             \Omega(\theta+\theta''(r_1,r_2,r_Y)), & \text{otherwise}.\\
                                                                                         \end{cases}
  \end{multline}
  
  The quantity $r_{Y, \text{deg}}'(r_1,\theta)$ is given by
  \begin{equation}
   r_{Y,\text{deg}}'(r_1,\theta) =  \sqrt{(r_1\cos(\theta))^2-r_1^2+nD} - r_1\cos(\theta)
  \end{equation}
  and $\theta''(r_1,r_2,r_Y)$ by
 \begin{align}
 \theta''(r_1,r_2,r_Y) = \begin{cases}
\vspace{0.1cm} 
	\arccos\left(\frac{r_1^2+r_Y^2-nD}{2 \cdot r_1\cdot r_Y}\right), & r_Y \leq \sqrt{r_1^2+nD}\\
\vspace{0.1cm}    
	\arccos\left(\frac{r_2^2+r_Y^2-nD}{2 \cdot r_2\cdot r_Y}\right), & r_Y \geq \sqrt{r_2^2+nD}\\
\vspace{0.1cm}																			
	\arcsin\left(\frac{\sqrt{nD}}{r_Y}\right), & \text{otherwise}. \\
\end{cases}
\end{align}
  
\end{prop}

\textit{Proof:} See Appendix~\ref{sec:proof_prop_2}.\hfill$\blacksquare$


 \begin{theorem}[Finite Blocklength Achievability]\label{theo:achievability}
  Let $R = R_G + R_S$, where $R_G$ and $R_S$ denote the rates of the employed gain and shape quantizers. Further, assume that
  the shape quantizer is a spherical code that has a guaranteed covering angle $\theta$ at rate $R_S$. At rate $R$, the achieved error probability is upper bounded by

  \begin{equation}  
   \Prob{\MAYBE} \leq \sum_{k\in[1:2^{nR_G}]} \int\limits_{r_{k-1}}^{r_k} \!f_{\norm{\bX}}(r_X)\id{r_X} \cdot \Pr\left\{\bY \in \Gamma^D(\CAP_{r_{k-1},r_k}(\hat{\bs}, \theta))\right\}.\label{eq:res_achievability}
  \end{equation}

 \end{theorem}
 
 \textit{Proof:} Theorem~\ref{theo:achievability} directly follows from Proposition~\ref{prop:prop_thick_cap} and the explanations leading to 
 \eqref{eq:prob_maybe}.\hfill$\blacksquare$
 
Averaging expression \eqref{eq:prob_y_thin_cap} with respect to $r$ will turn out to become a vital quantity to 
evaluate any signature assignment scheme $Q(\bX)$ based on a shape-gain quantizer, as it allows to evaluate the
effect of the shape quantization alone. This is due to the fact that by averaging over all $r$ we assume a genie-aided 
scenario where the decoder knows $\norm{\bX}$ exactly.
 
\vspace{2\baselineskip}
\begin{theorem}[Genie-Aided Finite Blocklength Achievability]\label{theo:achievability_genie}
  If the above setting is employed and a genie-aided knowledge about the exact value of the gain is available at the decoder,
  the probability of the query function returning $\MAYBE$ is upper bounded by
  
 \begin{equation}
 \Prob{\MAYBE} \leq \int\limits_{0}^\infty\!\Pr\left\{\bY \in \Gamma^D(\CAP_{r_X}(\hat{\bs}, \theta))\right\}f_{\norm{\bX}}(r_X)\id{r_X}.
 \end{equation}

 \end{theorem}
 
 \textit{Proof:} Theorem~\ref{theo:achievability_genie} directly follows from Proposition~\ref{prop:prop_thin_cap} and the 
 introductory explanations to this theorem.\hfill $\blacksquare$

Theorem~\ref{theo:achievability_genie}, while not pertaining to a directly implementable scheme, gives a bound on how much we 
can expect the bound \eqref{eq:res_achievability} to improve by employing the best possible scalar quantizer for this 
scenario.
 
 
 \subsection{Numerical Evaluation of the Integrals}
 \label{sec:num_eval1}
 
 As a prerequisite of Theorem 5, we assume the existence of a spherical code $\cC_S$ that provides a guarantee on the covering
 angle $\theta$ at a given rate $R_S$. As pointed out in Section~\ref{sec:sphere_packing_lattice}, we can use Dumer's 
 non-constructive achievability result on the covering density for spherical codes in \cite[Theorem 1]{Dumer07} for $n\geq4$, in order 
 to relate a given rate $R_S$ to a covering angle. Combining \eqref{eq:def_density_spherical_code} and \eqref{eq:bound_dumer}
 we can establish the following relation:
 
 \begin{equation}
  R_S = \frac{1}{n}\log_2\left(\frac{\vartheta(\cS^n)}{\Omega(\theta)}\right).\label{eq:rate_by_dumer}
 \end{equation}

 The overall rate is given as $R = R_S + R_G$, where the rate allocation is performed such that for a given $R_S$, 
 we search for the best $R_G$ within a discrete set of reasonable values.
 
 \begin{center}
 \includegraphics[scale=0.2]{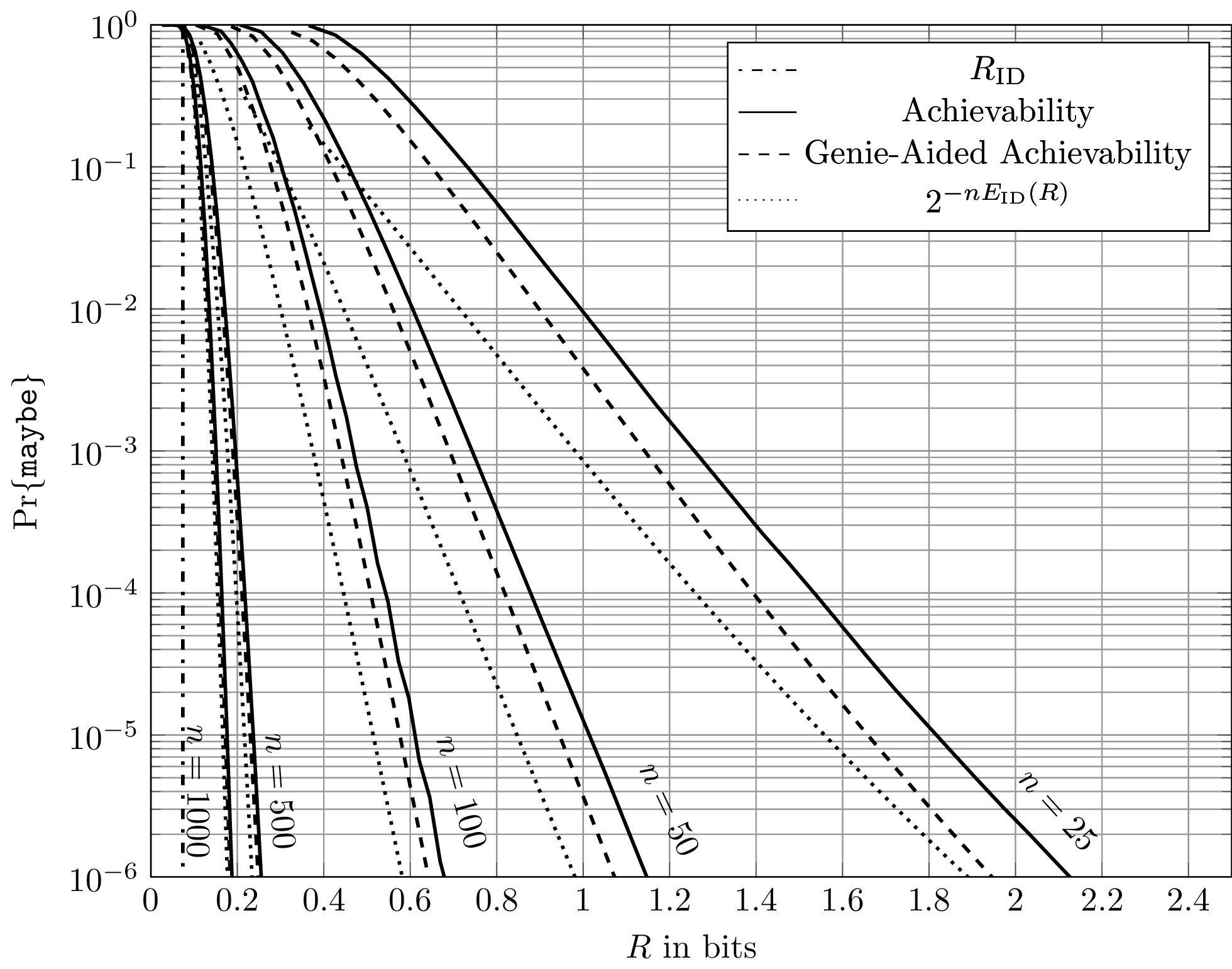}
 \captionof{figure}{Numerical evaluation of the achievability based on Theorem~\ref{theo:achievability} and the guaranteed covering angle of the non-constructive spherical code of \cite{Dumer07} with $D=0.1$.}
 \label{fig:res_achievability}
 \end{center}
 
 In Figure~\ref{fig:res_achievability}, the solid curves depict the numerical evaluation of the 3D-integral of \eqref{eq:res_achievability} for
 different dimensions $n$ and a desired similarity threshold of $D = 0.1$. Besides, we added the dotted curves that can be obtained by
 using the identification exponent $\bE_\ID(R)$ of \cite[Theorem 2]{ingberSQCD}. As expected, we see a convergence for  
 both curves for increasing $n$, as $\bE_\ID(R)$ was derived for the asymptotic case of infinite blocklengths. 
 Surprisingly, a good approximation can already be achieved for relatively small blocklengths of $n=500$ or $n=1000$.

 Another remark should be spent on the comparison of the solid and dashed curves of the genie-aided scheme imposed by 
 Theorem~\ref{theo:achievability_genie}: As pointed out before, the genied-aided curves depict the best performance which 
 can be hoped for with an optimal gain quantizer within our shape-gain quantization framework and provide an impression how 
 much would be gained if such a perfect gain quantizer were found. Since the gap between our MSE-cost criterion based  scalar
 quantizer and the genie-aided curve is negligible, we stick to the Lloyd-Max approach.

\section{Converse}
\label{sec:lb}

\subsection{Derivation of A Lower Bound}

Beyond the general achievability that has been shown in the previous section, we are interested in a converse that provides
a lower bound to the probability of $\MAYBE$. The derivations in this section closely follow the spirit of the
converse in \cite[Section IV.C]{ingberSQCD}, but put special emphasis on the details of the involved optimization.
Theorem~\ref{theo:lb} summarizes the main result.

\begin{theorem}\label{theo:lb}
 Let $(Q,g)$ be a rate $R$ compression scheme for a similarity threshold $D$. For $\eta >0$, define 
 $D' \triangleq \left(\sqrt{D}+\sqrt{1-\eta}-1\right)^2$ and $D''\triangleq \left(\sqrt{D'}+\sqrt{1-\eta}-1\right)^2$.
 Then, for any such $\eta$ that ensures $D > D' > D''$, we have the following lower bound on $\Prob{\MAYBE}$:
  
  \begin{align}
  	\Prob{\MAYBE} \geq  &\max_{c,\Omega^*,\eta} \quad c\cdot\Omega^*\cdot
 	\int_{\sqrt{n(1-\eta)}}^{\sqrt{n(1+\eta)}} f_{\norm{X}}(r_X)\id{r_X}\cdot
 	\int_{\sqrt{n(1-\eta)}}^{\sqrt{n(1+\eta)}} f_{\norm{Y}}(r_Y)\id{r_Y} \nonumber\\
 	&\text{s.t.}\quad 0 < c < 1\\
 	&\hphantom{s.t}\quad 0 < \Omega^* < 1\\
 	&\hphantom{s.t.}\quad \Omega\left(\theta_{D''}+\Omega^{-1}(p^*)\right) = \Omega^*\\
 	&\hphantom{s.t.}\quad R \leq \frac{1}{n}\log_2\left(\frac{1-c}{p^*}\right) 	
 \end{align}

 where  
 \begin{align}
   \theta_{D^{''}} &\triangleq \arccos\left(\frac{2-D''}{2}\right).
  \end{align}
 
 \end{theorem}
 \textit{Proof: } See Appendix~\ref{sec:proof_converse}.\hfill$\blacksquare$

 \subsection{Numerical Evaluation}
 
 Figure~\ref{fig:lower_bound} depicts the numerical results of Theorem~\ref{theo:lb} (dotted curves) and thereby allows for a comparison 
 to the achievability results (solid curves) already presented. Obviously, both bounds are not tight, but
 clearly become closer with increasing blocklength: For a blocklength of $n=25$ and $\Prob{\MAYBE} = 10^{-5}$, the corresponding
 rate lies within the interval of $\left[0.75;1.8\right]$, whereas the interval shrinks to $\left[0.3;0.6\right]$ for the blocklength
 $n=100$.
 
 \begin{center}
 \input{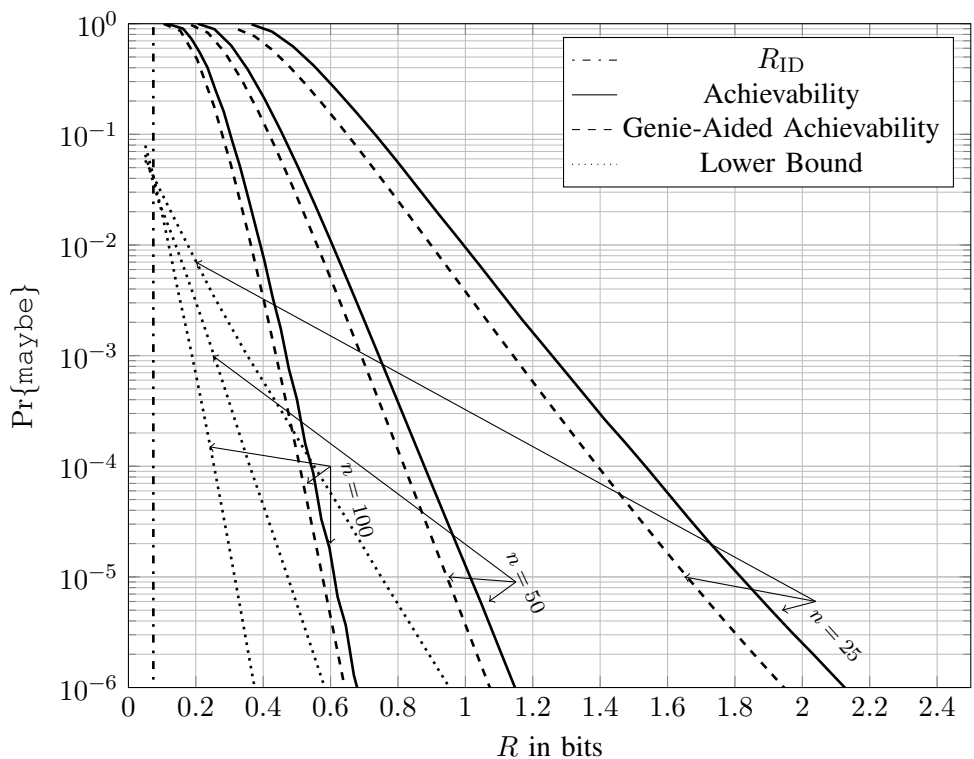}
 \captionof{figure}{Comparison of the achievability and the converse for a desired similarity level of $D=0.1$.}
 \label{fig:lower_bound}
 \end{center}

\section{Practical Implementation Of A Spherical Code}
\label{sec:practical_implementation}
\tikzstyle{block} = [draw, fill=white!20, rectangle, 
    minimum height=3em, minimum width=6em]
\tikzstyle{sum} = [draw, fill=white!20, circle, node distance=1cm]
\tikzstyle{input} = [coordinate]
\tikzstyle{output} = [coordinate]
\tikzstyle{pinstyle} = [pin edge={to-,thin,black}]

\subsection{Introduction}

The proposed achievability scheme in Sec.~\ref{sec:achievability} relies on the existence of a spherical code with an upper bound
on the covering angle. However, the previous numerical results in Figure~\ref{fig:res_achievability}
were all based on a non-constructive covering density result by Dumer \eqref{eq:bound_dumer} which provided little insight into how such a
spherical code can actually be implemented practically. For this purpose, we look for implementable spherical codes, such as those that
were introduced in 1996 by Hamkins in his PhD thesis \cite{hamkinsPhd} and subsequent work \cite{hamkinsWrappedCodes} \cite{hamkinsGaussSource}. In this work, he surveys several different approaches
based on lattices and establishes their optimality\footnote{Optimality hereby refers to the fact that
the covering density of the spherical code approaches the covering density of the underlying lattice if $d_\Lambda$ approaches zero.}
in an asymptotic sense.

For our purpose, we focus on wrapped spherical codes that are based on the fact that for any given dimension $n$ we can 
find a lattice~$\Lambda$ in~$\mathds{R}^{n-1}$  which is then ``wrapped'' onto a spherical shell~$\cS^n$ in $\Reals^n$ to obtain the 
codepoints. The properties of the underlying lattice, particularly its covering density~\eqref{eq:def_density_minimal}, then 
determine the performance of the shape quantization. For each dimension $n$, usually several different lattices can be found
that exhibit different covering densities. One of the densest lattices is given for $n=24$ and known as the \textit{Leech Lattice}
(cf. Appendix~\ref{sec:leech_lattice}). Because of its excellent properties, it serves as our model lattice in
the following section. At the same time, we emphasize that all arguments hold irrespective of the dimension or employed lattice.

So far, wrapped spherical codes have primarily been used for Gaussian source coding. In \cite{hamkinsGaussSource}, Hamkins and Zeger put special emphasis on the asymptotic case for high rates (which is equivalent to a small covering
radius $r_\Lambda^\mathrm{cov}$) and could show that the distance between two points on the lattice will be preserved 
on the spherical shell for asymptotically small $d_\Lambda$ \cite[Lemma 4.2]{hamkinsPhd}. 

In order for a scheme to be practical, we require that both the encoding stage (the mapping $Q(\cdot)$) and the decision functions be implementable. In light of this, any spherical code that is also implementable serves as a candidate for a practical compression-for-queries scheme. The non-trivial part is the decision function $g(\cdot)$. In Sec. \ref{sec:achievability}, we used a spherical code with a global guarantee for the covering angle. Therefore the scheme could be admissible since the expanded quantization cells are contained in expanded spherical caps, which are highly structured. Wrapped spherical codes, while implementable rather easily, do not come with a guarantee on the resulting covering angle.
For this purpose, we also provide a new derivation of an upper bound on the covering angle $\angle(\bS,\hat{\bS})$ of any wrapped spherical code, a result that may be of independent interest.

\subsection{Shape Quantization}
The implementation of the spherical code that is used for the shape quantization uses a wrapped spherical code and consists of three main steps. First, 
the input vector, which resides on an $n$-dimensional shell of the unit sphere~$\cS^n$ in ~$\Reals^n$, is mapped to the Euclidean 
space in $\Reals^{n-1}$ by using an appropriate mapping function $h(\cdot)$. Second, the obtained point is quantized to the nearest 
point of a lattice~$\Lambda$ in~$\Reals^{n-1}$ using a nearest neighbor search algorithm. Lastly, the 
quantized point is mapped back onto the spherical shell in $\Reals^n$ by using an inverse mapping function $h^{-1}(\cdot)$ which
results in the quantized output vector shown in Figure~\ref{fig:block_diagram}.

\vspace{\baselineskip}
\begin{tikzpicture}[auto,>=latex']
	\small{
		\node [input, name=input] {};
		\node [block, right of=input, node distance=2.6cm] (mapping) {$h(\cdot)$};
		\draw [draw,->] (input) -- node {$\bs \in \mathcal{S}^{n}$} (mapping);}
    \node [block, right of=mapping, node distance=4.6cm] (nns) {$Q_{\text{NN}}(\cdot)$};
		\draw [->] (mapping) -- node {$h_i(\bs) \in \cB^{n-1}$} (nns);		
		\node [block, right of=nns, node distance=5.5cm] (inv_mapping) {$h^{-1}(\cdot)$};
		\draw [->] (nns) -- node {$Q_{\text{NN}}(h(\bs)) \in \cB^{n-1}$} (inv_mapping);
		\node [output, right of=inv_mapping, node distance=2.6cm] (output) {};
		\draw [draw,->] (inv_mapping) -- node {$\hat{\bs} \in \mathcal{S}^{n}$} (output);
	\end{tikzpicture}
	
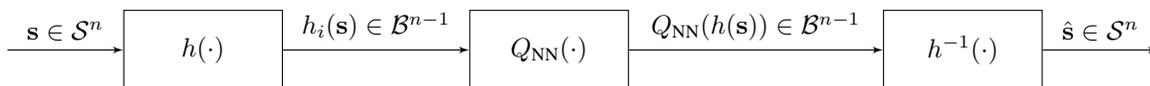
\captionof{figure}{Block diagram for the shape quantization process of a random vector $\bs$ using a wrapped spherical code. $h(\cdot)$ denotes the mapping function, $Q_{\text{NN}}(\cdot)$ is the lattice quantizer and $h^{-1}(\cdot)$ is the inverse mapping.}
	\label{fig:block_diagram}
\vspace{\baselineskip}

As a prerequisite, we partition the surface of $\cS^n$ in several regions which are called \textit{annuli}. Further, let 
the latitude of a point $\bs = (s_1,s_2,\ldots,s_n)^T\in\mathcal{S}^n$ be defined as 
$\arcsin(s_n)$ and $-\pi/2 = \alpha_0 < \ldots < 0 < \ldots < \alpha_N = \pi/2$ is a sequence of latitudes, where 
$N = \ceil{\frac{\pi}{\sqrt{d_\Lambda}}}+\left(\ceil{\frac{\pi}{\sqrt{d_\Lambda}}} \pmod{2}\right)$ and $\alpha_i = \pi(\frac{i}{N}-\frac{1}{2})$. The modulo operation is used to ensure that one annulus boundary has latitude $0$.
Then, the $i$-th annulus is defined as

\begin{align} \label{eq:def_annulus}
 A_i = \left\{\bs\in\mathcal{S}^n: \alpha_i \leq \arcsin(s_n) < \alpha_{i+1}\right\},
\end{align}

for $0 \leq i < N$.

\vspace{\baselineskip}
\subsubsection{Mapping Function}\label{sec:mapping_function}

The function $h(\cdot)$ is defined in a different manner for each annulus. For $\bs\in A_i \subseteq \mathcal{S}^n$, the mapping, now denoted $h_i(s)$, is defined as  


\begin{align}\label{eq:def_hi}
h_i(\bs) = 
\frac{\bs'}{\norm{\bs'}}\left(\norm{P_i(\bs)'} - \norm{P_i(\bs)-\bs}\right)_{+},
\end{align}
where $(\bx)_+\equiv \max(\mathbf{0},\bx)$ and the prime notation simply means the deletion of the last coordinate, i.e. $\bs'=(s_1,\ldots,s_{n-1})$. The point $P_i(\bs)$ is given as the solution to the optimization problem

\begin{align}\label{eq:def_pi}
	P_i(\bs) = \underset{\bz}{\arg\min} \norm{\bs-\bz}, 
	\quad\text{s.t. } \bz = 
		\begin{cases} 
		(z_1,z_2,\ldots,z_n = \sin(\alpha_i))\in\mathcal{S}^n & \text{if }\alpha_i \geq 0 \\
		(z_1,z_2,\ldots,z_n = \sin(\alpha_{i+1}))\in\mathcal{S}^n & \text{if }\alpha_i < 0 \\
		\end{cases}
\end{align}

\noindent \textit{Remark 1:} For facilitating the calculation of the upper bound on the angle, we further adapt the mapping
in \eqref{eq:def_hi} such that

\begin{align*}
h_i(\bs) = \mathbf{0}, \qquad \text{if}\quad & \arcsin(s_n) \geq \alpha_{N-1} + 2\arcsin\left(\frac{\cos(\alpha_{N-1})}{2}\right),\\ 
                                             & \arcsin(s_n) \leq \alpha_1-2\arcsin\left(\frac{\cos(\alpha_{1})}{2}\right).
\end{align*}
The above condition holds for vectors close to the north or south pole of a spherical shell.

\noindent \textit{Remark 2:} Analyzing the mapping, it becomes obvious that the image of each annulus under $h_i(A_i)$ is given by a
thick spherical shell. We exploit this property to count the number of possible codewords with the help of the theta function
\eqref{eq:def_lattice_theta2} in Section~\ref{sec:shape_rate}.

\vspace{\baselineskip}
\subsubsection{Lattice Quantizer}\label{sec:lattice_quant}
The lattice quantizer $Q_{\text{NN}}(\cdot)$ returns the lattice point which has the shortest distance to the mapped point 
$h_i(\bS)$ and is therefore implemented as a nearest neighbor quantizer. 
This step can naively be seen as an integer least-squares (ILS) problems: Mathematically speaking we assume a
point $h_i(\bs) \in \Reals^{n-1}$  and solve the optimization problem

\begin{align}
	\underset{\bb \in \Integers^{n-1}}\min \norm{h_i(\bs) - \bM\bb}^2,
\end{align}

where the generator matrix of the lattice is $\bM \in \Reals^{(n-1)\times (n-1)}$. Having gained insight into special
geometric properties of $\bM$, other approaches than ILS are usually prefered, as they yield solutions in a more efficient
manner.

\vspace{\baselineskip}
\begin{center}
\begin{tikzpicture}[scale=7]
 \draw (0,0) -- (1.2,0);
 \draw[name path=circle1] (0:1cm) arc (0:90:1cm);
 \draw[name path=annulus1] (0,0.5) -- (0.87,0.5);
 \draw[name path=annulus2] (0,0.866) -- (0.5,0.866);
 \coordinate (PiS) at (30:1cm);
 \path[name path=circle_big_help] (PiS) circle (.4cm);
\path[name path=circle_small_help] (PiS) circle (.2cm);
\coordinate (Pi1S) at (0.5,0.866);
 \draw [name intersections={of=circle1 and circle_big_help, by=S}] (S) node[right] {$\bs$};
\fill (S) circle [radius=.25pt];
 \draw[name path=PiSpath] (0,0) -- (PiS) node[right] {$P_i(\bs)$};
\draw (0,0) +(0:.1cm) arc (0:30:.1cm) node[right,yshift=-4pt] {$\alpha_i$};
\fill (PiS) circle [radius=.25pt];
 \draw[name path=Pi1Spath] (0,0) -- (Pi1S);
\draw (0,0) +(0:.2cm) arc (0:60:.2cm) node[right] {$\alpha_{i+1}$};
\draw (PiS) -- (S);
 \path [name intersections={of = annulus1 and circle_big_help}];
 \coordinate (P2) at (intersection-1);
\getxx{P2}
 \draw (P2) -- (\mydimm, 0) node[right,below,xshift=-10pt] {$h_i(\bs)$};
\fill (\mydimm,0) circle [radius=.25pt];
\draw (PiS) ++(130:0.4cm) arc (130:180:0.4cm);
 \path [name intersections={of = annulus1 and circle_small_help}];
 \coordinate (P1) at (intersection-1);
 \getx{P1}
 \draw (P1) -- (\mydim,0) node[right,below,xshift=1pt] {$Q_{\text{NN}}(h_i(\bs))$};
\fill (\mydim,0) circle [radius=.25pt];
\draw (PiS) ++(125:0.2cm) arc (125:180:0.2cm);
 \draw [name intersections={of=circle1 and circle_small_help, by=Shat}] (Shat) node[right,xshift=4pt,yshift=-4pt] {$\hat{\bs}$};
 \fill (Shat) circle [radius=.25pt];
\end{tikzpicture}
\captionof{figure}{Mapping, quantization and inverse mapping of a random vector $\bS$.}\label{fig:shape_quant}
\end{center}

\noindent \textit{Remark 3: }The process of the lattice quantization may return a point outside of the image of the original annulus. This is 
principally no problem for the algorithm since we keep the original annulus $i$ still in mind. However, in case the returned 
point is in the outside of $\cB^{n-1}$, i.e. if $\norm{Q_\mathrm{NN}(h_i(\bs))} > 1$, we scale the quantized vector back to the shell of the unit sphere. This step is necessary to ensure that the conditions under which the inverse mapping in Sec. \ref{sec:inv_mapping} is derived still hold.

\vspace{\baselineskip}
\subsubsection{Inverse Mapping} \label{sec:inv_mapping}

The inverse mapping $h_i^{-1}(\cdot)$ of a point in $\mathcal{B}^{n-1}$ back to the spherical shell~$\cS^n$ is 
performed as follows. Let $h_i(\bs)=\by$, then

\begin{align}\label{eq:norm_Y}
	\norm{\by} = \cos(\alpha_i) - 2\sin{\left(\frac{\abs{\arcsin{(s_n)}-\alpha_i}}{2}\right)}.
\end{align}

By reordering the previous equation we obtain

\begin{align}
	\arcsin{(s_n)} = \alpha_i \pm 2\arcsin{\left(\frac{\cos{\alpha_i}-\norm{\by}}{2}\right)},\label{eq:xn}
\end{align}
where the $\pm$-operator corresponds to the northern and southern hemisphere, respectively.

We use \eqref{eq:xn} and $\norm{\bs'}=\sqrt{1-s_{n}^2}$ in the original mapping function \eqref{eq:def_hi} to come up with
\begin{align}
	\bs' &=  \by \cdot \frac{\norm{\bs'}}{\norm{\by}}\\
			&= \by \cdot \left(\frac{\cos\left(\alpha_i \pm 2\arcsin{\left(\frac{\cos{\alpha_i}-\norm{\by}}{2}\right)}\right)}{\norm{\by}}\right)
\end{align}

and can finally state the inverse mapped point by adding the last coordinate $s_n$ \eqref{eq:xn} to $\bs'$ such that

\begin{align}\label{eq:add_last_coordinate}
	\bs = (\bs', s_n)^T.
\end{align}

The mapping, quantization and inverse mapping are illustrated graphically in Figure~\ref{fig:shape_quant}.

\noindent \textit{Remark 4: }The derivation of the inverse mapping $h_i^{-1}(\cdot)$ is based on the mapping function $h_i(\cdot)$ and the assumption that there is no quantization. If we consider cases where $\norm{\by}>h_i(P_i(\bs))$ (due to the quantization process), we observe a negative argument in the arcsine function of equation \eqref{eq:xn}. However, all formulas remain valid since the resulting point in equation \eqref{eq:add_last_coordinate} lies in the original (extended) annulus.


\vspace{\baselineskip}
\subsubsection{Encoding Process}
\label{sec:encoding_process}

Based on its quantized gain, annulus and index within its annulus, a vector is assigned an integer value which allows a unique encoding process.
In view of this we need to know in advance how many possible codepoints exist in one unit shell and how many possible codepoints there are in each annulus.
Having obtained the amount of codepoints within a unit shell and the gain partition of the original vector allows us to determine the number of codepoints in the unit shells with a lower index. This integer value then represents the quantized gain. 
Knowing how many codepoints there are in each annulus is necessary to determine all codepoints lying in the annuli with a lower index than the original annulus. Adding that part to the index of the codepoint within its own annulus represents the quantized shape.
Then, adding both integer values of quantized gain and shape will define the message. From this message, one can easily learn both the annulus index and the codepoint number.
 
\noindent \textit{Remark 5: }The counting of codepoints within a lattice is done by using the theta function which is described in detail in Sec. \ref{sec:shape_rate}.

%

\subsection{Maximum Covering Angle}
\label{sec:max_cov_angle}

If the previously described scheme is employed for designing a spherical code, Theorem~\ref{th:cov_angle} will state that the angle
between any vector on the shell of the unit sphere and its quantized version is always upper bounded by an angle $\theta$ such that the concept
of wrapped spherical codes provides a constructive method for our proposed achievability scheme of Section~\ref{sec:analysis_scheme}.

\vspace{\baselineskip}
\begin{theorem}[Upper Bound on the Covering Angle]\label{th:cov_angle}
The maximum covering angle $\theta$ between a vector $\bs\in A_i\subset\cS^n$ and its quantized version $\hat{\bs}\in\mathcal{S}^n$ constructed by wrapped spherical codes as described before is bounded for any dimension by

\begin{align}\label{eq:cov_angle}
	\theta \leq
	\begin{cases}
	\left(\frac{\pi}{2}\right) - 2\arcsin{\left(\frac{\cos(\alpha_i)-r_\Lambda^\mathrm{cov}}{2}\right)} & \text{if } Q_{\text{NN}}\left(h_i(\bs)\right)=\mathbf{0}\\
	2\cdot \arcsin\left(\frac{r_\Lambda^\mathrm{cov}}{2 \cdot \norm{h_i(\hat{\bs})}}\right) + 2\arcsin\left(\frac{\norm{\hat{\bs}-P_i(\hat{\bs})}+ r_\Lambda^\mathrm{cov}}{2}\right) - 2\arcsin\left(\frac{\norm{\hat{\bs}-P_i(\hat{\bs})}}{2}\right) & \text{else}.
	\end{cases}
\end{align}
\end{theorem}

\textit{Proof:} See Appendix~\ref{sec:proof_cov_angle}.

\begin{flushright}
  $\blacksquare$
\end{flushright}

We stress that the obtained upper bound is solely dependent on the vector $\hat{\bs}$ and the index $i$ of the annulus 
and $r_\Lambda^\mathrm{cov}$. Those quantities are both available at the decoder, which then uses this knowledge in order 
to detect similarity, i.e. to know whether the query sequence $\by$ is in the expanded spherical cap.

\noindent \textit{Remark 6: }As mentioned in remark 3, a vector that is quantized in the outside of $\cB^{n-1}$ requires 
a projection to the unit sphere. However, the derived bound on the covering angle still holds, since the scaling of that 
vector does not change the angle between original and (scaled) quantized vector. Furthermore, if $\hat{\bs}$ is outside 
$h_i(A_i)$, the bound holds due to the properties of the lattice (the maximum distance from the original vector 
to its quantized version is $r_\Lambda^\mathrm{cov}$).

\subsection{Numerical Performance Analysis}
\label{sec:performance_wrapped_codes}

\subsubsection{Introductory Remarks}
\label{sec:shape_rate}

In the following, we repeat the numerical simulations of Figure~\ref{fig:res_achievability} and compare those for dimension
$n=25$ to an achievability scheme that is based on actually implementable spherical codes. For that purpose, some further 
explanations are necessary: The following performance analysis employs a Semi-Monte-Carlo simulation, where we first 
generate standard Gaussian vectors of dimension $n=25$ and quantize the gain (Lloyd-Max) and shape (wrapped spherical code). 

As before, we have to address the issue of an optimal rate allocation, as it is generally not known beforehand, how much of
a given rate $R$ should be allocated to the shape or gain quantization. Optimally, the following optimization problem has to be
solved:

\begin{align*}
	\underset{R_G, R_S}\min \Prob{\MAYBE} \quad\text{s.t. }R=R_G+R_S.
\end{align*}

Since this problem is computationally cumbersome, we stick to the same method as in Section~\ref{sec:num_eval1} and
search for the best $R_G$ within a discrete set of reasonable values for a fixed $R_S$. At this point, we have not yet addressed 
the issue of how to obtain a spherical code for a desired rate $R_S$. As was described in the code construction, the performance of the wrapped
spherical code depends on the geometrical properties of the underlying lattice, in particular its covering radius $r_\Lambda^\mathrm{cov}$.

In view of this, we make use of another property of a lattice, namely its theta function, which counts the number of lattice points 
on a spherical shell with a given discrete radius. A common way to define this property is by expressing it 
as a complex polynomial in $q = \mathrm{e}^{j\pi z}$. For our counting purposes, one particular form of the theta function
is especially insightful: If $N_m$ denotes the number of lattice points $\bx \in \Lambda$ with 
$\norm{\bx}^2 = m$, then the theta function can be written as

\begin{align}
 \Theta_\Lambda(z) \triangleq \sum_{m = 0}^\infty N_m q^m.\label{eq:def_lattice_theta2}
\end{align}

Thus, the polynomial coefficients $N_m$ of the theta function can be used for counting the number of lattice points lying on
a spherical shell $\cS^n_{\sqrt{m}}$. In order to relate this number of points to the actual number of codepoints of a wrapped spherical code,
we have to take into account that the image of an annulus $A_i \subset \cS^n$ under $h_i(\cdot)$ is a thick spherical shell
$\cS_{\tilde r_-,\tilde r_+}$ \eqref{eq:def_shell}. If additionally the nearest neighbor quantization is considered, we see that
all codewords for any random vector $\bS\in A_i$ must also lie within a thick shell $\cS_{r_-, r_+}$ where $r_-$ and $r_+$ are defined as in algorithm \ref{algo:calc_shape_rate}.
Evidently, due to the discrete nature of the coefficients of the theta function, also the rate of the shape
quantizer is of discrete nature. By scaling the lattice accordingly (cf. Section~\ref{sec:lattices}), we can therefore 
indirectly adjust the rate as well. 

Recalling that the points on the boundaries of an annulus $A_i$, $P_i(\bs)$, are mapped to $\Reals^{n-1}$ by simply deleting 
their last coordinate, we arrive at $\norm{P_i(\bs)'} = \cos(\alpha_i)$ and can summarize the entire procedure in
Algorithm~\ref{algo:calc_shape_rate}.

\vspace{\baselineskip}
\begin{algorithm}[H]\label{algo:calc_shape_rate}
 \KwData{desired lattice scaling (here defined by $r_\Lambda^\mathrm{cov}$) number of annuli $N$, set of angles $\alpha_i$}
 \SetKwFunction{CountCodepoints}{CountCodepoints}
 $M = 0$\\
 \For{$i$ = $1$ to $N$}{
    $r_- \leftarrow \max\left(0,\cos(\alpha_i)-2\sin(\pi/(2N))-r_\Lambda^\mathrm{cov}\right)$\\
    $r_+ \leftarrow \min\left(1+r_\Lambda^\mathrm{cov}, \cos(\alpha_i)+r_\Lambda^\mathrm{cov}\right)$\\
	  $M = M + \CountCodepoints(\Lambda, r_-, r_+)$\\
 }
 $R_S = \frac{1}{n}\log_2(M)$
 \caption{Relating a given lattice scaling to the rate $R_S$ of the shape quantizer.}
\end{algorithm}
\vspace{\baselineskip}

Algorithm~\ref{algo:calc_shape_rate} refers to the function \texttt{CountCodepoints()} which basically implements the theta
function of the lattice and returns $\sum_{m: r_- \leq \sqrt{m} \leq r_+} N_m$, i.e. the sum of all codepoints within the 
specified thick spherical shell. 

\subsubsection{Results of the Comparison}

Figure~\ref{fig:semi_mc} exhibits a comparison of the non-constructive result discussed earlier with the implementable
scheme presented in this section. For this purpose, we have conducted a Monte-Carlo simulation that evaluates 
$\Prob{\MAYBE}$ based on 1000 samples for the input vector. 

 The computation is facilitated by first conditioning on the value of $\bX$, i.e.
 $\Prob{\MAYBE} = \mathrm{E}\left[\Prob{\MAYBE|\bX}\right]$, where the expectation is calculated with Monte-Carlo. The 
 inner expression is evaluated as follows: for each $\bx$, we quantize its shape using the wrapped spherical code and 
 obtain a bound to the covering angle according to Theorem~\ref{th:cov_angle} and also keep the real angle for reference. 
 With the limits on the gain of $\bx$ (obtained using the gain quantization), we invoke Proposition~\ref{prop:prop_thick_cap}
 and calculate the conditional probability $\Prob{\MAYBE|\bX}$.
 
In addition to the new achievability and genie-aided achievability curves that are based
on the upper bound of the covering angle given by Theorem~\ref{th:cov_angle}, we include two further curves that use the
actual angle $\angle(\bS,\hat{\bS})$ so that the quality of the upper bound can be assessed as well.

We observe that the practical implementation achieves a performance that  -- for a given $\Prob{\MAYBE}$ -- is only approx. 1 Bit worse than the 
non-constructive achievability results from Sec. \ref{sec:achievability}. Further, having perfect knowledge about the covering angle, the performance gap of 
the proposed scheme compared to the non-constructive achievability decreases steadily with higher rates 
and is smaller than $0.2$ Bit for $R>2$ Bit.

\begin{center}
%
%
\begin{tikzpicture}

\begin{axis}[%
width=4.40888888888889in,
height=3.47733333333333in,
scale only axis,
xmin=0,
xmax=3,
xlabel={$R$ in bits},
xmajorgrids,
ymode=log,
ymin=1e-06,
ymax=1,
yminorticks=true,
ylabel={$\mathrm{Pr}\{\texttt{maybe}\}$},
ymajorgrids,
yminorgrids,
legend style={
at={(0,0)},
anchor=south west},
legend entries={\tiny $R_\mathrm{ID}$,
                \tiny Achievability (Non-Constructive),
                \tiny Genie-Aided Achievability (Non-Constructive),
                \tiny Achievability (Scheme: Upper Bound),
                \tiny Genie-Aided Achievability (Scheme: Upper Bound),
                \tiny Achievability (Scheme: Real Angle)
                }
]
\addlegendimage{black,dash pattern=on 1pt off 3pt on 3pt off 3pt}
\addlegendimage{black,solid}
\addlegendimage{black,dashed}
\addlegendimage{black,solid,mark=square,mark options = {solid}}
\addlegendimage{black,dashed,mark=square,mark options = {solid}}
\addlegendimage{black,dotted,mark=square,mark options = {solid}}
\addplot [
color=black,
dashed,
line width=1.0pt,
forget plot
]
table[row sep=crcr]{
0.325268358567763 0.888002666973173\\
0.375268358567763 0.767544440686604\\
0.425268358567763 0.585402303171354\\
0.475268358567763 0.417539567846752\\
0.525268358567763 0.285419398076075\\
0.575268358567763 0.189556650076297\\
0.625268358567763 0.123357728346132\\
0.675268358567763 0.079114818185489\\
0.725268358567763 0.0502094051961918\\
0.775268358567763 0.0316272182157188\\
0.825268358567763 0.019819453611165\\
0.875268358567763 0.0123785340097405\\
0.925268358567763 0.00771670725508908\\
0.975268358567763 0.00480731475958985\\
1.02526835856776 0.00299584802338396\\
1.07526835856776 0.00186918990066475\\
1.12526835856776 0.00116847901967933\\
1.17526835856776 0.00073231316383739\\
1.22526835856776 0.000460384876245729\\
1.27526835856776 0.00029047228281655\\
1.32526835856776 0.000184007163963847\\
1.37526835856776 0.000117078597667272\\
1.42526835856776 7.48482386478087e-05\\
1.47526835856776 4.80928239592271e-05\\
1.52526835856776 3.10665501775068e-05\\
1.57526835856776 2.01802819472917e-05\\
1.62526835856776 1.31849782714119e-05\\
1.67526835856776 8.66634747545495e-06\\
1.72526835856776 5.73158588071595e-06\\
1.77526835856776 3.8147375769646e-06\\
1.82526835856776 2.55544799998176e-06\\
1.87526835856776 1.72320143806324e-06\\
1.92526835856776 1.16981976690732e-06\\
1.97526835856776 7.99570255789552e-07\\
2.02526835856776 5.50280152908676e-07\\
2.07526835856776 3.81355152576369e-07\\
2.12526835856776 2.66144472093851e-07\\
2.17526835856776 1.87053407473188e-07\\
2.22526835856776 1.32399830869794e-07\\
2.27526835856776 9.43824829195508e-08\\
2.32526835856776 6.77612663357421e-08\\
2.37526835856776 4.89956395269399e-08\\
2.42526835856776 3.56791645431751e-08\\
2.47526835856776 2.61664526624466e-08\\
2.52526835856776 1.93257261243436e-08\\
2.57526835856776 1.4373810668786e-08\\
2.62526835856776 1.07654982717266e-08\\
2.67526835856776 8.11895808114644e-09\\
2.72526835856776 6.16517310869693e-09\\
2.77526835856776 4.71347435178435e-09\\
2.82526835856776 3.62791751506122e-09\\
2.87526835856776 2.81100174732613e-09\\
2.92526835856776 2.19238649597608e-09\\
2.97526835856776 1.72102900676916e-09\\
3.02526835856776 1.35967534518308e-09\\
3.07526835856776 1.08097933499969e-09\\
3.12526835856776 8.6475348876814e-10\\
3.17526835856776 6.96011239191129e-10\\
3.22526835856776 5.63565222907852e-10\\
3.27526835856776 4.5901834641414e-10\\
3.32526835856776 3.76033742737398e-10\\
3.37526835856776 3.09803770496231e-10\\
3.42526835856776 2.56661792892567e-10\\
3.47526835856776 2.1379689538078e-10\\
3.52526835856776 1.79043189474683e-10\\
3.57526835856776 1.50723426942761e-10\\
3.62526835856776 1.27532354105874e-10\\
3.67526835856776 1.0844928554282e-10\\
3.72526835856776 9.2672264402815e-11\\
3.77526835856776 7.95682456256015e-11\\
3.82526835856776 6.8635227705751e-11\\
3.87526835856776 5.9473335580391e-11\\
3.92526835856776 5.17626394947018e-11\\
3.97526835856776 4.52460654821413e-11\\
4.02526835856776 3.97161714107998e-11\\
4.07526835856776 3.50048704630392e-11\\
4.12526835856776 3.09754115476274e-11\\
4.17526835856776 2.75160951419756e-11\\
4.22526835856776 2.45353290573081e-11\\
4.27526835856776 2.19577229436611e-11\\
4.32526835856776 1.97209912541766e-11\\
4.37526835856776 1.77734878996071e-11\\
4.42526835856776 1.60722363701798e-11\\
4.47526835856776 1.45813499443093e-11\\
4.52526835856776 1.32707601547315e-11\\
4.57526835856776 1.21151897342089e-11\\
4.62526835856776 1.10933201507675e-11\\
4.67526835856776 1.01871145660318e-11\\
4.72526835856776 9.38126536040757e-12\\
4.77526835856776 8.66274183159754e-12\\
4.82526835856776 8.02041871645539e-12\\
4.87526835856776 7.44477013563219e-12\\
4.92526835856776 6.92761666359132e-12\\
4.97526835856776 6.46191567279851e-12\\
};
\addplot [
color=black,
solid,
line width=1.0pt,
forget plot
]
table[row sep=crcr]{
0.365268358567763 0.98789686257956\\
0.426986766765118 0.846065250990814\\
0.488705174962472 0.621759772398838\\
0.550423583159826 0.416905238600429\\
0.612141991357181 0.262103458042393\\
0.673860399554535 0.16099277834077\\
0.735578807751889 0.0969984584756673\\
0.797297215949243 0.0571060803292507\\
0.859015624146598 0.0329179250980063\\
0.920734032343952 0.0190121183133181\\
0.982452440541306 0.0111328526399508\\
1.04417084873866 0.00646281046711247\\
1.10588925693601 0.00371926314862204\\
1.16760766513337 0.00214043444371969\\
1.22932607333072 0.00127266005757654\\
1.29104448152808 0.000755431772368402\\
1.35276288972543 0.000447613120758403\\
1.41448129792279 0.00026465249384848\\
1.47619970612014 0.000162578324214143\\
1.53791811431749 9.67627449509872e-05\\
1.59963652251485 5.75382973728963e-05\\
1.6613549307122 3.4193882615026e-05\\
1.72307333890956 2.06557465110445e-05\\
1.78479174710691 1.28106453697683e-05\\
1.84651015530427 7.97517473046415e-06\\
1.90822856350162 4.98405972578496e-06\\
1.96994697169897 3.14306931862802e-06\\
2.03166537989633 1.98096673478892e-06\\
2.09338378809368 1.24821671667536e-06\\
2.15510219629104 7.88932008455343e-07\\
2.21682060448839 5.12416328167624e-07\\
2.27853901268575 3.37483725708431e-07\\
2.3402574208831 2.23587663834459e-07\\
2.40197582908045 1.49036108998222e-07\\
2.46369423727781 1.01183648778666e-07\\
2.52541264547516 6.96288424093763e-08\\
2.58713105367252 4.78326487131507e-08\\
2.64884946186987 3.35184483611713e-08\\
2.71056787006723 2.36749478400602e-08\\
2.77228627826458 1.68614012436281e-08\\
2.83400468646194 1.21590168608309e-08\\
2.89572309465929 8.97413932747977e-09\\
2.95744150285664 6.55404735341692e-09\\
3.019159911054 4.74204398964988e-09\\
3.08087831925135 3.45238858637819e-09\\
3.14259672744871 2.54433910467793e-09\\
3.20431513564606 1.90571401087021e-09\\
3.26603354384342 1.43672990225157e-09\\
3.32775195204077 1.09028444628755e-09\\
3.38947036023812 8.33977773502547e-10\\
3.45118876843548 6.48984858440065e-10\\
3.51290717663283 5.08352614858199e-10\\
3.57462558483019 4.00831826729461e-10\\
3.63634399302754 3.18155258134207e-10\\
3.6980624012249 2.56128048334994e-10\\
3.75978080942225 2.07809306955962e-10\\
3.8214992176196 1.69694431702849e-10\\
3.88321762581696 1.38342203653958e-10\\
3.94493603401431 1.13770188576174e-10\\
4.00665444221167 9.41000392785819e-11\\
4.06837285040902 7.82742857885561e-11\\
4.13009125860637 6.55399017480595e-11\\
4.19180966680373 5.54379905731841e-11\\
4.25352807500108 4.71404778521123e-11\\
4.31524648319844 4.02938021726546e-11\\
4.37696489139579 3.46188308790994e-11\\
4.43868329959315 3.00049884399701e-11\\
4.5004017077905 2.61409848861987e-11\\
4.56212011598786 2.2881269090775e-11\\
4.62383852418521 2.01203105949281e-11\\
4.68555693238256 1.78071802094638e-11\\
4.74727534057992 1.58461402821288e-11\\
4.80899374877727 1.41586493467842e-11\\
4.87071215697463 1.27023297394934e-11\\
4.93243056517198 1.14506927984146e-11\\
4.99414897336934 1.03763064305795e-11\\
5.05586738156669 9.43691697285066e-12\\
5.11758578976404 8.61304442507833e-12\\
5.1793041979614 7.88964814884581e-12\\
5.24102260615875 8.96129228263453e-12\\
};
\addplot [
color=black,
dash pattern=on 1pt off 3pt on 3pt off 3pt,
line width=1.0pt,
forget plot
]
table[row sep=crcr]{
0.0740005814437767 3.11569931964159e-11\\
0.0740005814437767 1\\
};
\addplot [
color=black,
solid,
line width=1.0pt,
forget plot,
mark=square,
mark repeat=10,
mark options = {solid}
]
table[row sep=crcr]{
1.44919104370294 0.690325068024734\\
1.46829058484986 0.657097565975442\\
1.48739012599678 0.62387006392615\\
1.50648966714371 0.590642561876858\\
1.52558920829063 0.557415059827566\\
1.54468874943755 0.524187557778275\\
1.56378829058448 0.402150668334298\\
1.5828878317314 0.358709348528675\\
1.60198737287833 0.315268028723052\\
1.62108691402525 0.271826708917429\\
1.64018645517217 0.310988303218556\\
1.6592859963191 0.270388825528228\\
1.67838553746602 0.2297893478379\\
1.69748507861294 0.237873026753742\\
1.71658461975987 0.174949121604668\\
1.73568416090679 0.153678574063913\\
1.75478370205371 0.132408026523159\\
1.77388324320064 0.111137478982405\\
1.79298278434756 0.114822831280751\\
1.81208232549448 0.0985846783191981\\
1.83118186664141 0.0823465253576453\\
1.85028140778833 0.0661083723960924\\
1.86938094893525 0.068226754157206\\
1.88848049008218 0.0572178395697386\\
1.9075800312291 0.0462089249822711\\
1.92667957237602 0.0328876527157384\\
1.94577911352295 0.0295962842456593\\
1.96487865466987 0.0263049157755802\\
1.98397819581679 0.01932517205426\\
2.00307773696372 0.0164325701671819\\
2.02217727811064 0.0140778670734325\\
2.04127681925756 0.0129225150736272\\
2.06037636040449 0.0117671630738218\\
2.07947590155141 0.00969344910315469\\
2.09857544269834 0.00740307161604162\\
2.11767498384526 0.00511590099374175\\
2.13677452499218 0.00438188314231408\\
2.15587406613911 0.00375399641916288\\
2.17497360728603 0.00345149150131994\\
2.19407314843295 0.00314898658347699\\
2.21317268957988 0.00284648166563405\\
2.2322722307268 0.00237931225218716\\
2.25137177187372 0.00176201786411783\\
2.27047131302065 0.00137424719093906\\
2.28957085416757 0.00110932165338255\\
2.30867039531449 0.000964477260876171\\
2.32776993646142 0.000838060110674229\\
2.34686947760834 0.000711642960472284\\
2.36596901875526 0.000614715581762546\\
2.38506855990219 0.00052597075972991\\
2.40416810104911 0.000437225937697272\\
2.42326764219603 0.000370780255941242\\
2.44236718334296 0.000312450952030936\\
2.46146672448988 0.000255961850666456\\
2.4805662656368 0.000214651098271661\\
2.49966580678373 0.000173340345876868\\
2.51876534793065 0.000143207332515462\\
2.53786488907757 0.000124468646605841\\
2.5569644302245 0.000103696079917005\\
2.57606397137142 8.69912747866006e-05\\
2.59516351251834 7.65811889648679e-05\\
2.61426305366527 6.60151652102488e-05\\
2.63336259481219 5.58730062829405e-05\\
2.65246213595911 4.95257994873004e-05\\
2.67156167710604 4.31785926916603e-05\\
2.69066121825296 3.6688985577453e-05\\
2.70976075939989 3.01406589539256e-05\\
2.72886030054681 2.35923323303983e-05\\
2.74795984169373 1.85209882335031e-05\\
2.76705938284066 1.65053231000006e-05\\
2.78615892398758 1.30080521139725e-05\\
2.8052584651345 1.02949575065573e-05\\
2.82435800628143 8.92752696259758e-06\\
2.84345754742835 7.56009641863782e-06\\
2.86255708857527 6.19266587467807e-06\\
2.8816566297222 5.16140982625272e-06\\
2.90075617086912 4.56060731884494e-06\\
2.91985571201604 3.95980481143716e-06\\
2.93895525316297 3.35900230402938e-06\\
2.95805479430989 2.7581997966216e-06\\
};
\addplot [
color=black,
dashed,
line width=1.0pt,
forget plot,
mark=square,
mark repeat=10,
mark options = {solid}
]
table[row sep=crcr]{
1.44919104370294 0.571331581977662\\
1.46829058484986 0.54102260068712\\
1.48739012599678 0.510713619396578\\
1.50648966714371 0.480404638106036\\
1.52558920829063 0.450095656815494\\
1.54468874943755 0.419786675524952\\
1.56378829058448 0.314284609991238\\
1.5828878317314 0.278248992781418\\
1.60198737287833 0.242213375571598\\
1.62108691402525 0.206177758361778\\
1.64018645517217 0.266863838920143\\
1.6592859963191 0.23292475698827\\
1.67838553746602 0.196521427128716\\
1.69748507861294 0.178472469323585\\
1.71658461975987 0.128239869422283\\
1.73568416090679 0.111919399801292\\
1.75478370205371 0.0955989301803007\\
1.77388324320064 0.0792784605593097\\
1.79298278434756 0.0820594008044833\\
1.81208232549448 0.0699577657180072\\
1.83118186664141 0.0578561306315311\\
1.85028140778833 0.045754495545055\\
1.86938094893525 0.0471968012891301\\
1.88848049008218 0.0391958683208359\\
1.9075800312291 0.0311949353525418\\
1.92667957237602 0.0217631080690574\\
1.94577911352295 0.0194728980992299\\
1.96487865466987 0.0171826881294023\\
1.98397819581679 0.0124312902926364\\
2.00307773696372 0.0104620203686292\\
2.02217727811064 0.00886759814608436\\
2.04127681925756 0.00810897255738044\\
2.06037636040449 0.00735034696867652\\
2.07947590155141 0.00598316296865349\\
2.09857544269834 0.00447240550898963\\
2.11767498384526 0.00302468374189994\\
2.13677452499218 0.00256610853388546\\
2.15587406613911 0.002175161834282\\
2.17497360728603 0.00199155379690166\\
2.19407314843295 0.00180794575952133\\
2.21317268957988 0.001624337722141\\
2.2322722307268 0.00134297433398093\\
2.25137177187372 0.000972487182369657\\
2.27047131302065 0.000743433448384822\\
2.28957085416757 0.00059007737108058\\
2.30867039531449 0.000508473457148172\\
2.32776993646142 0.000437880385090804\\
2.34686947760834 0.000367287313033433\\
2.36596901875526 0.000313870398294835\\
2.38506855990219 0.000265219369273351\\
2.40416810104911 0.000216568340251866\\
2.42326764219603 0.000181121153432256\\
2.44236718334296 0.000150479863656046\\
2.46146672448988 0.000120881914281896\\
2.4805662656368 9.9889639730087e-05\\
2.49966580678373 7.88973651782788e-05\\
2.51876534793065 6.38316796802503e-05\\
2.53786488907757 5.48074213571834e-05\\
2.5569644302245 4.57831630341168e-05\\
2.57606397137142 3.67589047110497e-05\\
2.59516351251834 3.20405125388127e-05\\
2.61426305366527 2.79713785171738e-05\\
2.63336259481219 2.39022444955348e-05\\
2.65246213595911 1.98331104738958e-05\\
2.67156167710604 1.68088818606362e-05\\
2.69066121825296 1.39723309739741e-05\\
2.70976075939989 1.11357800873119e-05\\
2.72886030054681 8.29922920064977e-06\\
2.74795984169373 6.14297442289075e-06\\
2.76705938284066 5.39416114925928e-06\\
2.78615892398758 4.6453478756278e-06\\
2.8052584651345 3.89653460199632e-06\\
2.82435800628143 3.21976644795012e-06\\
2.84345754742835 2.81897557849462e-06\\
2.86255708857527 2.41818470903912e-06\\
2.8816566297222 2.01739383958361e-06\\
2.90075617086912 1.61660297012811e-06\\
2.91985571201604 2.21796869162103e-06\\
2.93895525316297 1.85660878569119e-06\\
2.95805479430989 1.49524887976134e-06\\
};
\addplot [
color=black,
dotted,
line width=1.0pt,
forget plot,
mark=square,
mark repeat=10,
mark options = {solid}
]
table[row sep=crcr]{
1.44919104370294 0.00931333903589146\\
1.46829058484986 0.00843630606984435\\
1.48739012599678 0.00755927310379724\\
1.50648966714371 0.00668224013775013\\
1.52558920829063 0.00580520717170302\\
1.54468874943755 0.00492817420565591\\
1.56378829058448 0.00286090320268899\\
1.5828878317314 0.00233790357005241\\
1.60198737287833 0.00181490393741583\\
1.62108691402525 0.00129190430477925\\
1.64018645517217 0.00113931337366209\\
1.6592859963191 0.000924764354178939\\
1.67838553746602 0.000710215334695788\\
1.69748507861294 0.000675357028200017\\
1.71658461975987 0.000285074237219719\\
1.73568416090679 0.000277186873719441\\
1.75478370205371 0.000269299510219162\\
1.77388324320064 0.000261412146718883\\
1.79298278434756 0.000184101303773707\\
1.81208232549448 0.000145859978773545\\
1.83118186664141 0.000107618653773383\\
1.85028140778833 6.93773287732218e-05\\
1.86938094893525 6.67774892084128e-05\\
1.88848049008218 5.30974556664365e-05\\
1.9075800312291 3.94174221244601e-05\\
1.92667957237602 2.5876850672734e-05\\
1.94577911352295 2.28048794561568e-05\\
1.96487865466987 1.97329082395796e-05\\
1.98397819581679 1.56507417487767e-05\\
2.00307773696372 1.31395858365676e-05\\
2.02217727811064 1.11988605755691e-05\\
2.04127681925756 8.78279109889201e-06\\
2.06037636040449 7.01353616810012e-06\\
2.07947590155141 5.97120103697201e-06\\
2.09857544269834 5.54529697839294e-06\\
2.11767498384526 4.25723786008602e-06\\
2.13677452499218 3.77262129972232e-06\\
2.15587406613911 3.34873132192128e-06\\
2.17497360728603 2.29607901643225e-06\\
2.19407314843295 1.91858922001455e-06\\
2.21317268957988 1.65602360237451e-06\\
2.2322722307268 1.51923847369096e-06\\
2.25137177187372 1.38245334500742e-06\\
2.27047131302065 1.24566821632388e-06\\
2.28957085416757 1.02877714048162e-06\\
2.30867039531449 7.85834013938079e-07\\
2.32776993646142 6.54820038447134e-07\\
2.34686947760834 5.4453682296066e-07\\
2.36596901875526 4.88107245401102e-07\\
2.38506855990219 4.31677667841547e-07\\
2.40416810104911 3.75469771056515e-07\\
2.42326764219603 3.345485073292e-07\\
2.44236718334296 2.93627243601886e-07\\
2.46146672448988 2.52705979874571e-07\\
2.4805662656368 2.22031038686407e-07\\
2.49966580678373 1.91731602540006e-07\\
2.51876534793065 1.63771099424147e-07\\
2.53786488907757 1.40358435285914e-07\\
2.5569644302245 1.16945771147682e-07\\
2.57606397137142 1.04207575979065e-07\\
2.59516351251834 9.54838746292807e-08\\
2.61426305366527 8.67601732794967e-08\\
2.63336259481219 7.82882180852245e-08\\
2.65246213595911 7.2070199872607e-08\\
2.67156167710604 6.58521816599896e-08\\
2.69066121825296 5.96341634473722e-08\\
2.70976075939989 5.3486324603714e-08\\
2.72886030054681 4.81540299429353e-08\\
2.74795984169373 4.28217352821565e-08\\
2.76705938284066 3.74894406213778e-08\\
2.78615892398758 3.2157145960599e-08\\
2.8052584651345 2.84918301432818e-08\\
2.82435800628143 2.61504316901402e-08\\
2.84345754742835 2.38090332369986e-08\\
2.86255708857527 2.1467634783857e-08\\
2.8816566297222 1.96093977316784e-08\\
2.90075617086912 1.83698231262062e-08\\
2.91985571201604 1.7130248520734e-08\\
2.93895525316297 1.58906739152618e-08\\
2.95805479430989 1.46510993097896e-08\\
};
\end{axis}
\end{tikzpicture}

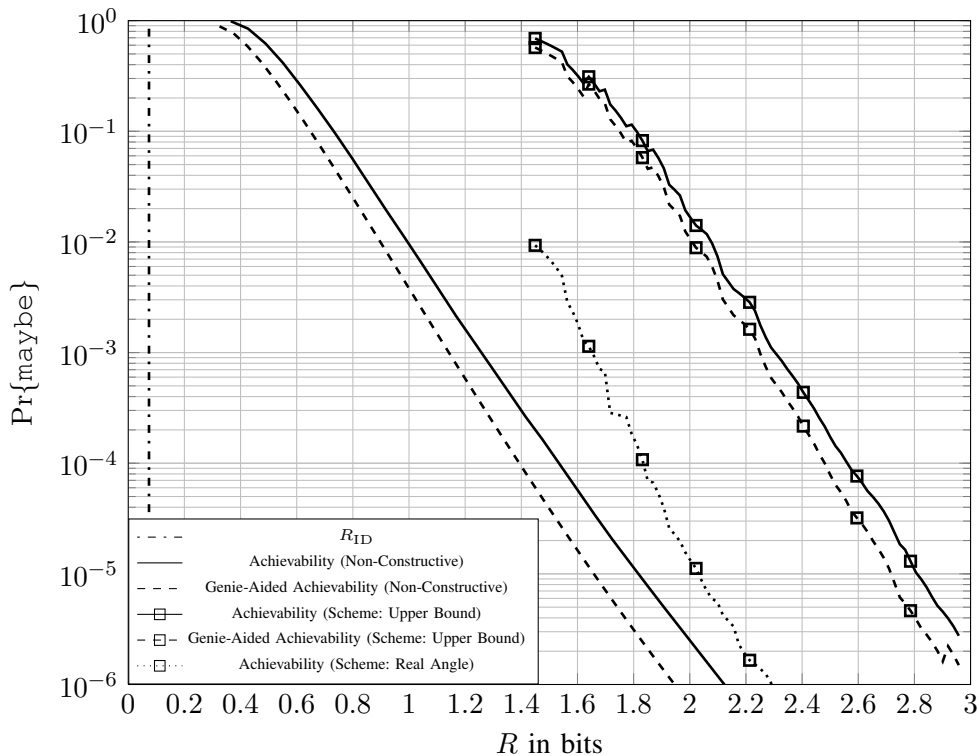
\captionof{figure}{Numerical evaluation of the performance of the scheme using wrapped spherical %
codes (based on the Leech lattice) in $n=25$ compared to the non-constructive achievability result from Section~\ref{sec:achievability}}%
\label{fig:semi_mc}
\end{center}

\section{Conclusion}
\label{sec:remarks}

We studied the rate-reliability trade-off for the finite blocklength regime, when similarity queries with Gaussian input data
and a quadratic similarity measure are performed. We provided expressions that allow for a numerically computable characterization of
a lower and upper bound on the error probability and also compared it to formulas for the asymptotic case.

Throughout the paper, we emphasized the possibility of applying the derived framework to practical problems. As a matter
of fact, we concentrated on implementable spherical codes, namely those based on wrapped spherical codes, and proved an upper bound 
on the covering angle so that this spherical code can indeed be employed for the shape quantization.

To sum up, our practical scheme can always be used if the following conditions are met: First, the \textit{covering radius 
$r_{\Lambda}$} for the desired lattice in $\Reals^n$ - which upper bounds the distance between any arbitrary point 
in this dimension and a lattice point - has to be known. Second, an efficient \textit{nearest neighbour search} must exist that 
guarantees the nearest neighbour in the lattice can be found. Since our bound on the covering angle is a function of the covering radius of the underlying lattice, the above mentioned two points are all that is needed to perform similarity queries with our scheme.

While the method described in the paper is implementable, as opposed to previous work, we do not yet have any optimality 
guarantees regarding the performance of the scheme (such as, for example, the optimality of shape-gain quantizers for source 
coding at high rates \cite{hamkinsGaussSource}). This is left for future work. 


Another direction for future work is testing the usefulness of our schemes in real life scenarios. Applications where queries would be performed in a feature domain, where the Gaussian assumption is natural, seem particularly promising although the robustness property of the Gaussian distribution \cite[Sec. III-D]{ingberSQCD} suggests schemes that are designed assuming Gaussianity may be much more widely applicable.  

\appendices
\section{}
\label{sec:proof_prop_1}

\textit{Proof of Proposition~\ref{prop:prop_thin_cap}}: The cruical part for proving this proposition involves the probability 
density of the random variable $\bY|\norm{\bY} =r_Y$. Using \eqref{eq:def_chi} and \eqref{eq:def_area_sphere}, one obtains

\begin{align}
 f_{\bY|\norm{\bY}}(\by|r_Y) = \frac{(2\pi)^{-\frac{n}{2}}\mathrm{e}^{-r_Y^2/2}}{\frac{2^{1-\frac{n}{2}r^{n-1}}\mathrm{e}^{-r_Y^2/2}}{\Gamma\left(\frac{n}{2}\right)}} = \frac{1}{\abs{\cS^n_{r_Y}}}\label{eq:uniform_on_sphere}, 
\end{align}

which shows that a Gaussian random variable $\bY|\norm{\bY} =r_Y$ is uniformly distributed over a sphere with radius $r_Y$ in
$n$ dimensions. 

Consequently, for calculating the probability $\Prob{\bY \in \Gamma^D(\CAP_r(\hat{\bs}, \theta)) \left|\right. \norm{\bY} = r_Y}$
it only matters what fraction of the sphere $\cS^n_{r_Y}$ is included within in the $D$-expansion of the respective cap. 
Hence, we can make use of the Omega function described in \eqref{eq:def_omega} to calculate this particular fraction. As
noted before, this function only depends on the angle of corresponding cap. After all, the entire problem boils down to the calculation 
of the angle that is passed to the Omega function. Given a certain relation of the given scheme parameters and $r_Y$, several 
different cases have to be distinguished for that and are summarized for reference in the following:

  \begin{equation}
   \Prob{\bY \in \Gamma^D(\CAP_r(\hat{\bs}, \theta)) \left|\right. \norm{\bY} = r_Y} =\\ \begin{cases}
                                                                                             0, & |r-r_Y| > \sqrt{nD}\\
                                                                                             1, & r_Y \leq r_{Y, \text{deg}}(r,\theta) \\
                                                                                             \Omega(\theta+\theta'(r_Y,r)), & \text{otherwise}\label{eq:prob_y_thin_cap_cases}.\\
                                                                                          \end{cases}
  \end{equation}
  
The first case (cf. situations \circled{1a}, \circled{1b} in Figure~\ref{fig:prob_y_thin_cap}) can easily be determined: If $r_Y$ is too small 
or too large, $\bY$ can not lie inside the $D$-expansion of the thin cap. 

Regarding the third case (cf. situation \circled{3} in Figure~\ref{fig:prob_y_thin_cap}), $\bY$ may lie inside the $D$-expansion and we account for the possible fraction of $\cS^n_{r_Y}$ that
is part of this set by introducing the expansion angle $\theta'(r_Y,r)$. Applying the law of cosines to the triangle 
$(\mathbf{0},\bx,\by)$, one obtains

\begin{equation}
  \theta'(r_Y,r)\triangleq \arccos\left(\frac{r^2+r_Y^2-nD}{2\cdot r\cdot r_Y}\right).\label{eq:theta_prime_thin_cap}
\end{equation}

The second case  (cf. situation \circled{2} in Figure~\ref{fig:prob_y_thick_cap}) describes the degenerate situation, 
when the radius $r_Y$  is so small such that the sphere $\cS^n_{r_Y}$ is contained in the expanded cap. 
As Figure~\ref{fig:prob_y_thin_cap_deg} reveals, this is the case for $r_Y \leq r_{Y,\text{deg}}(r,\theta)$ as given in \eqref{eq:rYdegen}.

\begin{equation}
   r_{Y,\text{deg}}(r,\theta) =  \sqrt{(r\cos(\theta))^2-r^2+nD} - r\cos(\theta)
   \label{eq:rYdegen}
\end{equation}

\begin{center}
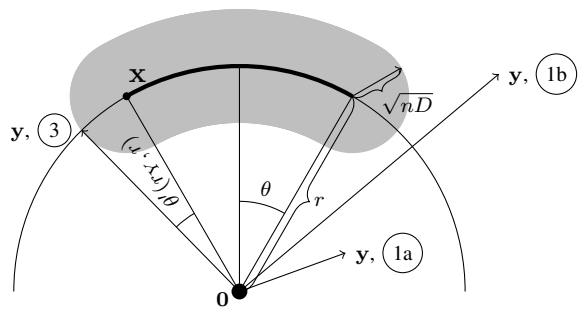

\begin{tikzpicture}[scale=3]

\foreach \a in {120,...,60}
    \filldraw[gray!50] (\a:1cm) circle (.25);

\draw (0:1) arc (0:180:1);

\draw[->] (60:1) -- +(30:0.25);
\draw[decorate,decoration={brace,amplitude=4pt,mirror}] (60:1) -- +(30:0.25) node[right,midway,xshift=-2pt,yshift=-8pt] {\scriptsize $\sqrt{nD}$};
\fill (0,0) circle[radius=1pt] node[left,yshift=-3pt] {\scriptsize $\mathbf{0}$};
\draw[ultra thick] (60:1cm) arc (60:120:1);
\draw (0,0) -- (60:1); 
\fill (120:1) circle[radius=.5pt] node[right,xshift=-3pt,yshift=8pt] {$\bx$};
\draw[decorate,decoration={brace,amplitude=6pt,mirror}] (0,0) --(60:1) node[right,midway,xshift=3pt,yshift=-3pt] {\scriptsize $r$};
\draw (0,0) -- (120:1);
\draw (0,0) -- (90:1);

\draw[->] (0,0) -- (134.36:1) node[left] {\scriptsize $\by$, \circled{3}};
\draw[->] (0,0) -- (20:.5) node[right] {\scriptsize $\by$, \circled{1a}};
\draw[->] (0,0) -- (40:1.5) node[right] {\scriptsize $\by$, \circled{1b}};

\draw (60:.4) arc (60:90:.4) node[xshift=10pt,yshift=-2pt,above] {\scriptsize $\theta$};
\draw (120:.4) arc (120:134.36:.4) node[right,rotate=127,xshift=3pt,yshift=-5pt] {\scriptsize $\theta'(r_Y,r)$};
\end{tikzpicture}
\captionof{figure}{Probability of $\bY$ falling into the $\Gamma^D$-expansion of a thin cap: Cases 1a, 1b and 3}
\label{fig:prob_y_thin_cap}
\end{center}

\begin{center}
\begin{tikzpicture}[scale=1.5]
\foreach \a in {120,...,60}
    \filldraw[gray!50] (\a:1) circle (1.5);

\draw (0:1) arc (0:180:1);

\draw[->] (60:1) -- +(30:1.5); 
\draw[decorate,decoration={brace,amplitude=4pt,mirror}] (60:1) -- +(30:1.5) node[right,midway,xshift=-5pt,yshift=-10pt] {\scriptsize $\sqrt{nD}$};

\fill (0,0) circle[radius=2pt] node[left,yshift=-3pt] {\scriptsize $\mathbf{0}$};
\draw[ultra thick] (60:1cm) arc (60:120:1);

\draw[decorate,decoration={brace,amplitude=4pt,mirror}] (0,0) --(60:1) node[right,midway,xshift=3pt,yshift=-3pt] {\scriptsize $r$};
\draw (0,0) -- (60:1);
\draw (0,0) -- (120:1);
\draw (0,0) -- (90:1);
\draw[->] (0,0) -- (30:0.5) node[right,xshift=-8pt,yshift=-8pt] {\scriptsize $\by$, \circled{2}};

\draw[->] (0,0) -- (0,-0.5482);
\draw[decorate,decoration={brace,amplitude=4pt}] (0,0) -- (0,-0.5482) node[right,midway,xshift=4pt,yshift=-3pt] {\scriptsize $r_{Y, \text{deg}}(r,\theta)$};

\draw (60:.4) arc (60:90:.4) node[xshift=5pt,above,yshift=-2pt] {\scriptsize $\theta$};
\end{tikzpicture}
\captionof{figure}{Probability of $\bY$ falling into the $\Gamma^D$-expansion of a thin cap: Degenerated case 2}
\label{fig:prob_y_thin_cap_deg}
\end{center}
 

\section{}
\label{sec:proof_prop_2}

\textit{Proof of Proposition~\ref{prop:prop_thick_cap}}: We follow the same argumentation as in the previous proof. 
The calculation of the probability of the random variable
$\bY|\norm{\bY} = r_Y$ falling into a thick cap, i.e. $\Prob{\bY \in \Gamma^D(\CAP_{r_1,r_2}(\hat{\bs}, \theta)) \left|\right. \norm{\bY} = r_Y}$,
boils down to the computation of the covered fraction of the sphere $\cS_{r_Y}^n$. However, the conditions 
for the different cases now have to be adapted:

\begin{align}
   \Prob{\bY \in \Gamma^D(\CAP_{r_1,r_2}(\hat{\bs}, \theta)) \left|\right. \norm{\bY} = r_Y} = \begin{cases}
											      0, & r_Y < r_1 - \sqrt{nD}\\
											         & r_Y > r_2 + \sqrt{nD} \\
                                                                                             1, & r_Y \leq r_{Y, \text{deg}}'(r_1, \theta) \\
                                                                                             \Omega(\theta+\theta''(r_1,r_2,r_Y)), & \text{otherwise}.\label{eq:prob_y_thick_cap_cases}\\
                                                                                         \end{cases}
\end{align}

Cases \circled{1a} ($r_Y < r_1 - \sqrt{nD}$) and \circled{1b} ($r_Y < r_2 - \sqrt{nD}$) (cf. Figure~\ref{fig:prob_y_thick_cap})
denote those situations when no part of the sphere $\cS^n_{r_Y}$ is included within the $\Gamma^D$-expansion of the thick cap. 

Case three (cf. \circled{3} in Figure~\ref{fig:prob_y_thick_cap}) turns out to be slightly more involved as in Proposition~\ref{prop:prop_thin_cap},
as the thickness of cap has be taken into account. Having said this, one is able to make the distinction between three additional regions which
are seperated by the boundaries $r_1' = \sqrt{r_1^2+nD}$ and $r_2' = \sqrt{r_2^2+nD}$. Those can be derived with the 
Pythagorean theorem for the respective triangles drawn in Figure~\ref{fig:prob_y_thick_cap}. Applying the law of cosines 
eventually to one of the appropriate triangles $(\mathbf{0},\bx_1,\by)$, $(\mathbf{0},\bx_2,\by)$ and $(\mathbf{0},\bx_3,\by)$
yields the following distinction for the expansion angle $\theta''(r_1,r_2,r_Y)$:

\begin{align}
 \theta''(r_1,r_2,r_Y) = \begin{cases}
\vspace{0.1cm} 
	\arccos\left(\frac{r_1^2+r_Y^2-nD}{2 \cdot r_1\cdot r_Y}\right), & r_Y \leq r_1' \\
\vspace{0.1cm}    
	\arccos\left(\frac{r_2^2+r_Y^2-nD}{2 \cdot r_2\cdot r_Y}\right), & r_Y \geq r_2' \\
\vspace{0.1cm}																			
	\arcsin\left(\frac{\sqrt{nD}}{r_Y}\right), & \text{otherwise}. \\
\end{cases}\label{eq:theta_r1r2}
\end{align}

Concerning the second case in \eqref{eq:prob_y_thick_cap_cases}, the remarks of Proposition~\ref{prop:prop_thin_cap} apply
analogously with the quantity $r_{Y, \text{deg}}'(r_1, \theta)$ being defined in the same 
way as in \eqref{eq:rYdegen} and $r$ replaced by $r_1$.


\vspace{2\baselineskip}

\begin{center}
\begin{tikzpicture}[scale=3]

\def\angle{120,...,60}
\def\radius{1,1.01,...,2}

\foreach \a in \angle {
	\foreach \r in \radius {
		\filldraw[gray!50] (\a:\r) circle (0.4cm);
		}
}

\fill (0,0) circle[radius=.5pt] node[left,yshift=-3pt] {\scriptsize $\mathbf{0}$};
\draw (0:1cm) arc (0:180:1cm);
\draw (0:2cm) arc(0:180:2cm);							
\draw[ultra thick] (60:1cm) arc (60:120:1cm);
\draw[ultra thick] (60:2cm) arc (60:120:2cm);
\draw (0,0) -- (60:2cm);
\draw (0,0) -- (120:2cm);
\draw (0,0) -- (90:2cm); 
\draw[decorate,decoration={brace,amplitude=6pt}] (0,0) --(90:1) node[left,midway,xshift=-3pt] {\scriptsize $r_1$};
\draw[decorate,decoration={brace,amplitude=6pt,mirror}] (0,0) --(90:2) node[right,midway,xshift=3pt,yshift=3pt] {\scriptsize $r_2$};

\fill (120:1) circle[radius=.5pt] node[right,xshift=-3pt,yshift=10pt] {$\bx_1$};
\fill (120:1.5) circle[radius=.5pt] node[right,xshift=-3pt,yshift=8pt] {$\bx_2$};
\fill (120:2) circle[radius=.5pt] node[right,xshift=-3pt,yshift=10pt] {$\bx_3$};

\draw (60:0.4cm) arc (60:90:0.4cm) node[above right,xshift=0.1cm] {\scriptsize $\theta$};

\draw (60-23.07:1.076) -- (60:1);
\draw (60-11.53:2.0396) -- (60:2);


\draw (0,0) -- (60-23.07:1.076);
\draw (0,0) -- (60-11.53:2.0396);

\draw[decorate,decoration={brace,amplitude=6pt,mirror}] (0,0) -- (60-23.07:1.076) node[right,midway,xshift=3pt,yshift=-2pt] {\scriptsize $r_1'$};
\draw[decorate,decoration={brace,amplitude=6pt,mirror}] (0,0) -- (60-11.53:2.0396) node[right,midway,xshift=5pt,yshift=2pt] {\scriptsize $r_2'$};

\draw (60:1.95cm) -- (58.5:1.95cm) -- (58.5:2.005cm);
\draw (60:0.95cm) -- (57:0.95cm) -- (57:1.005cm);

\draw (120:.4) arc (120:139.471:.4) node[right,rotate=127,xshift=-2pt,yshift=-5pt] {\scriptsize $\theta''(r_1,r_2,r_Y)$};

\draw[->] (0,0) -- (139.471:1.2) node[left] {\scriptsize $\by$, \circled{3}};
\draw[->] (0,0) -- (20:.5) node[right] {\scriptsize $\by$, \circled{1a}};
\draw[->] (0,0) -- (70:2.6) node[right] {\scriptsize $\by$, \circled{1b}};

\end{tikzpicture}

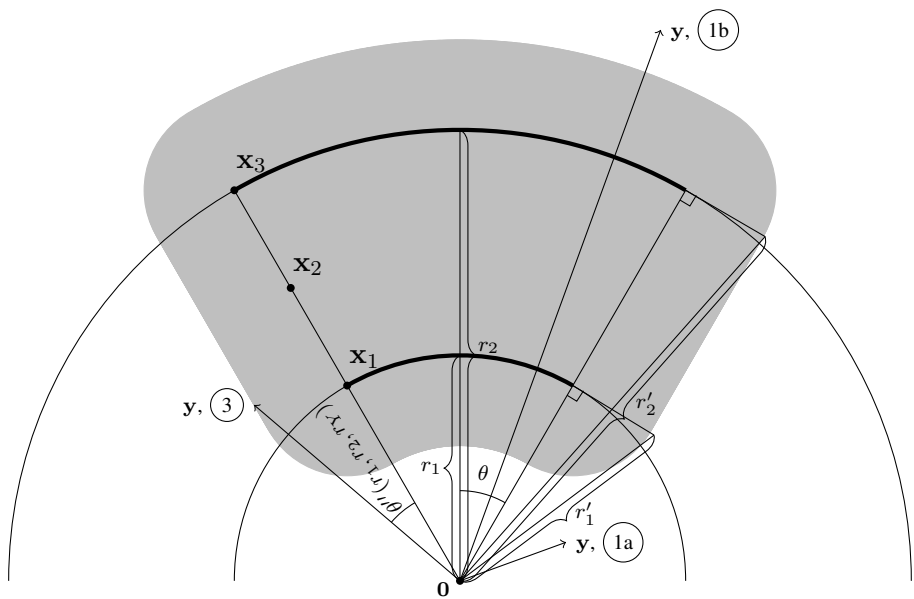
\captionof{figure}{Probability of $\bY$ falling into an expanded thick cap: Cases 1a, 1b and 3}
\label{fig:prob_y_thick_cap}
\end{center}

\section{}
\label{sec:leech_lattice}

The Leech Lattice \cite[Chapter 4.11]{ConwaySloane1993} is a well-known lattice in $n=24$ dimensions that is widely used in practice
as it provides the densest lattice covering in this dimension.

The Leech Lattice was chosen a model lattice for this paper, as most of its properties are explored in detail and its geometry exhibits
features that allow for convenient computations. It can be constructed in a large variety of ways, such as Golay codes and laminated
lattices. We rely on its generation via its generator matrix $\bM \in \Reals^{24\times24}$ that is given for reference in 
\cite[Chapter 4, Figure 4.12]{ConwaySloane1993}. Its determinant and hence the volume of the fundamental region evaluates as 
$\abs{\det(\bM)} = 1$. Further, the minimal distance is $d_{\Lambda_{24}}=2$ and relates directly to the covering radius as
$r_{\Lambda_{24}}^\mathrm{cov} = \frac{1}{2}d_{\Lambda_{24}} \sqrt{2} = \sqrt{2}$. Consequently, the packing and covering densities
are given as

\begin{align}
 \vartheta^\mathrm{cov}(\Lambda) &\triangleq \frac{V\left(\cS_{r_{\Lambda_{24}}^\mathrm{cov}}\right)}{\abs{\det(\bM)}} = \frac{\pi^{12}}{12!} \approx 0.001930,\label{eq:def_pack_cov_density}\\
 \vartheta^\mathrm{pack}(\Lambda)&\triangleq \frac{V\left(\cS_{r_{\Lambda_{24}}^\mathrm{pack}}\right)}{\abs{\det(\bM)}} = \frac{(2\pi)^{12}}{12!} \approx 7.9035\label{eq:def_lattice_cov_density}.
\end{align}

Its theta function is given by

\begin{align}
 \Theta_{\Lambda_{24}}(z) = \sum_{n = 0}^\infty N_m q^m = 1 + 196560q^4 + 16773120q^6 + 398034000q^8 + \ldots.
\end{align}


\section{}
\label{sec:proof_converse}

For proving Theorem~\ref{theo:lb} we start with inequality (94) of \cite{ingberSQCD} which reads as

\begin{align}\label{eq:converse_maybe_1}
  \Prob{\MAYBE} \geq  	\int_{\sqrt{n(1-\eta)}}^{\sqrt{n(1+\eta)}} f_{\norm{X}}(r_X)\id{r_X}\cdot
 	\int_{\sqrt{n(1-\eta)}}^{\sqrt{n(1+\eta)}} f_{\norm{Y}}(r_Y)\id{r_Y} \cdot\sum_{i=1}^{2^{nR}} p_i \cdot\Omega\left(\theta_{D''}+\Omega^{-1}(p_i)\right),
\end{align}
where its full derivation can be traced back in the aforementioned paper.

The set of probabilities $\left\{p_i\right\}_{i=1}^{2^{nR}}$ expresses the probability of a Gaussian variable
$\bX$ being an element of the set of all points that have been mapped to one of the $i \in \left[1:2^{nR}\right]$ possible signatures 
by a particular choice of the signature function $Q(\cdot)$ (cf. (73) in \cite{ingberSQCD}). By construction, the elements $p_i$ of the
respective set sum up to one.

At that point Lemma~4 of \cite{ingberSQCD} comes into play, which we state here for reference:

\begin{lemnonum}[Lemma~4, \cite{ingberSQCD}]\label{lem:converse}
    Let $0<\Omega^*<1$ and $0 < c < 1$ be given constants. Define $p^*$ to be the solution to $\Omega(\theta_{D''} + \Omega^{-1}(p))=\Omega^*$. Then if
    \begin{equation}\label{eqn:Sum_pi_Omega}
      \sum_{i=1}^{2^{nR}} p_i \cdot
    \Omega\left(\theta_{D''}+\Omega^{-1}(p_i)\right) \leq c\cdot \Omega^*,
    \end{equation}
    then
    \begin{equation}\label{eqn:Markov}
      R \geq \frac{1}{n}\log \frac{1-c}{p^*}.
    \end{equation}
\end{lemnonum}

This Lemma now becomes vital as it allows for a reformulation of \eqref{eq:converse_maybe_1}, when Lemma~4 is used the other way around:

\begin{align}
  \Prob{\MAYBE} \geq \int_{\sqrt{n(1-\eta)}}^{\sqrt{n(1+\eta)}} f_{\norm{X}}(r_X)\id{r_X}\cdot
 	\int_{\sqrt{n(1-\eta)}}^{\sqrt{n(1+\eta)}} f_{\norm{Y}}(r_Y)\id{r_Y} \cdot\ c\cdot \Omega^*,
\end{align}

if $R \leq \frac{1}{n}\log \frac{1-c}{p^*}$.

The best lower bound obviously can then be gained if we try to optimize the above expression with regard to $c$, $\Omega^*$ and $\eta$:

  \begin{align}
  	\Pr\{\texttt{maybe}\} \geq  &\max_{c,\Omega^*,\eta} \quad c\cdot\Omega^*\cdot
 	\int_{\sqrt{n(1-\eta)}}^{\sqrt{n(1+\eta)}} f_{\norm{X}}(r_X)\id{r_X}\cdot
 	\int_{\sqrt{n(1-\eta)}}^{\sqrt{n(1+\eta)}} f_{\norm{Y}}(r_Y)\id{r_Y}\\
 	&\text{s.t.}\quad 0 < c < 1\\
 	&\hphantom{s.t}\quad 0 < \Omega^* < 1\\
	&\hphantom{s.t}\quad 0 < \eta < 1\\
 	&\hphantom{s.t.}\quad \Omega\left(\theta_{D''}+\Omega^{-1}(p^*)\right) = \Omega^*\\
 	&\hphantom{s.t.}\quad R \leq \frac{1}{n}\log_2\left(\frac{1-c}{p^*}\right).
 \end{align}


\section{}
\label{sec:proof_cov_angle}


\textit{Proof of Theorem~\ref{th:cov_angle}}: First of all, we introduce an auxiliary variable $\hat{\hat{\bS}}$ and 
bound the covering angle with the help of the triangle inequality, such that

\begin{align}\label{eq:tri_ineq}
	\theta = \angle(\bX,\hat{\bX}) = \angle(\bS,\hat{\bS}) \leq \angle(\bS,\hat{\hat{\bS}}) + \angle(\hat{\hat{\bS}},\hat{\bS}).
\end{align}


\begin{tikzpicture}[scale=2.4]
 \draw (0,0) -- (4,0);
 \draw (0,0) -- (30:4cm);
 \draw (0:.25cm) arc (0:30:.25cm) node[right,yshift=-5pt] {$\delta$};
 \coordinate (O) at (0,0);
 \coordinate (Xhat) at (2,0);
 \coordinate (X) at (30:2.5cm);
 \coordinate (Xhathat) at (30:2cm);
 \coordinate (XLX) at (30:3cm);
 \coordinate (XLXhat) at (3,0);

 \draw (X) -- (Xhat) node[midway,right] {$\tilde r_{\Lambda}$};
 \draw (XLX) -- (XLXhat);
 \draw (Xhat) -- (Xhathat) node[midway,right] {$b$};
 \draw (Xhathat) -- (X) node[below,midway] {$a$};
 \fill (Xhat) circle[radius=.5pt] node[below] {$Q_{\text{NN}}(h_i(\bS))=h_i(\hat{\bS})$};
 \fill (X) circle[radius=.5pt] node[above] {$h_i(\bS)$};
 \fill (Xhathat) circle[radius=.5pt] node[above] {$h_i(\hat{\hat{\bS}})$};
 \fill (XLX) circle[radius=.5pt] node[above,xshift=-8pt,yshift=0pt] {$P_i'(\bS)$};
 \fill (XLXhat) circle[radius=.5pt] node[below] {$P_i'(\hat{\bS})$};
 \fill (O) circle[radius=.5pt] node[below] {$\mathbf{0}$};
\end{tikzpicture}
\captionof{figure}{Setting in $\Reals^{n-1}$ to derive a bound on $\angle{(\hat{\bS},\hat{\hat{\bS}})}$ in \eqref{eq:tri_ineq} for the maximum covering angle $\theta$.}\label{graph:theta_part1}

\vspace{\baselineskip}
Without loss of generality we can assume a setting in $\Reals^{n-1}$ as shown in Figure~\ref{graph:theta_part1},
where $\hat{\hat{\bS}}$ was chosen such that it is on the same line as $h_i(\bS)$ and $P_i'(\bS)$ and has distance
$\norm{h_i(\hat{\bS})}$ from the origin. In general $\tilde r_{\Lambda} $ is bounded by 
$\tilde r_{\Lambda} = \norm{h_i(\bS)-Q_{\text{NN}}(h_i(\bS))} \leq r_{\Lambda}$ because of the properties of the 
lattice $\Lambda$ and the implementation of $Q_{\text{NN}}(\cdot)$ as a nearest neighbor quantizer.

We argue that $\angle{(P_i(\bS),P_i(\hat{\bS})}) = \angle{(P_i(\hat{\hat{\bS}}),P_i(\hat{\bS})}) = \angle{(\hat{\hat{\bS}},\hat{\bS})}$ 
as follows. Since $h_i(\hat{\hat{\bS}})$ is on the same line as $h_i(\bS)$ and $P_i'(\bS)$ by construction,
the definition of \eqref{eq:def_pi} implies $P_i'(\bS) = P_i'(\hat{\hat{\bS}})$.

Due to the special property that the mapping for all points that lie exactly on the boundary between two annuli is just the 
deletion of the last coordinate, the distance between $P_i'(\bS)=P_i'(\hat{\hat{\bS}})$ and $P_i'(\hat{\bS})$ does not change 
when applying the inverse mapping, so that 
$\norm{P_i'(\bS) - P_i'(\hat{\bS})}=\norm{P_i(\bS) - P_i(\hat{\bS})}=\norm{P_i(\hat{\hat{\bS}}) - P_i(\hat{\bS})}$.

We have then found a quantity to describe an upper bound of the first part of \eqref{eq:tri_ineq} because 
$\hat{\hat{\bS}}$ and $\hat{\bS}$ have the same latitude and 
$\angle{(\hat{\hat{\bS}},\hat{\bS})}=\angle{(P_i(\hat{\hat{\bS}}),P_i(\hat{\bS}))}$.

More precise, the distance between $h_i(\hat{\hat{\bS}})$ and $h_i(\bS)$ can be calculated as
\begin{align}
	a &= \abs{\norm{h_i(\bS)}-\norm{h_i(\hat{\hat{\bS}}))}} \leq \abs{\tilde r_{\Lambda}+\norm{h_i(\hat{\bS})}-\norm{h_i(\hat{\hat{\bS}})}} = \tilde r_{\Lambda} \leq r_{\Lambda},
\end{align}
where the inequality follows from $\norm{h_i(\bS)}\leq \tilde r_{\Lambda}+\norm{h_i(\hat{\bS})}$.

Furthermore the distance $b$ between $h_i(\hat{\hat{\bS}})$ and $h_i(\hat{\bS})$ can be bounded as follows. We make use of 
the law of cosines and obtain

\begin{align}
	\tilde r_\Lambda^2 &= a^2 + b^2 - 2 a b \cos \left(\frac{\pi}{2} \pm \frac{\delta}{2}\right)\\
				& = a^2+b^2\left(\frac{1 \mp a}{\norm{h_i(\hat{\bS})}}\right).
\end{align}
The derivation is straightforward and by using $\norm{h_i(\hat{\bS})}\leq 1$ and $a\leq r_\Lambda < 1$ we can state
\begin{align}
	b &\leq \tilde r_{\Lambda}.
\end{align}

Eventually, the angle $\delta$ is upper bounded by 
\begin{align}
	\delta &= 2\cdot \arcsin{\frac{b/2}{\norm{h_i(\hat{\bS})}}} \leq 2\cdot \arcsin{\frac{\tilde r_{\Lambda}}{2 \cdot \norm{h_i(\hat{\bS})}}}\\
		&\leq 2\cdot \arcsin{\frac{r_{\Lambda}}{2 \cdot \norm{h_i(\hat{\bS})}}}.
\end{align}

\vspace{\baselineskip}
\begin{center}
\begin{tikzpicture}[scale=7]
 \draw (0,0) -- (1.2,0);
 \draw[name path=circle1] (0:1cm) arc (0:90:1cm);
 \draw[name path=annulus1] (0,0.5) -- (0.866,0.5);
 \draw[name path=annulus2] (0,0.866) -- (0.5,0.866);
 \coordinate (PiS) at (30:1cm);
 \path[name path=circle_big_help] (PiS) circle (.4cm);
\path[name path=circle_small_help] (PiS) circle (.2cm);
\coordinate (Pi1S) at (0.5,0.866);
 \draw [name intersections={of=circle1 and circle_big_help, by=S}] (S) node[right] {$\bS$};
\fill (S) circle [radius=.25pt];
 \draw[name path=PiSpath] (0,0) -- (PiS) node[right] {$P_i(\bS)=P_i(\hat{\hat{\bS}})$};
\draw (0,0) +(0:.1cm) arc (0:30:.1cm) node[right,yshift=-4pt] {$\alpha_i$};
\fill (PiS) circle [radius=.25pt];
 \draw[name path=Pi1Spath] (0,0) -- (Pi1S);
\draw (0,0) +(0:.2cm) arc (0:60:.2cm) node[right,xshift=17pt,yshift=-25pt] {$\alpha_{i+1}$};
\draw (PiS) -- (S);
 \path [name intersections={of = annulus1 and circle_big_help}];
 \coordinate (P2) at (intersection-1);
\getxx{P2}
 \draw (P2) -- (\mydimm, 0) node[right,below,xshift=-10pt,yshift=-5pt] {$h_i(\bS)$};
\fill (\mydimm,0) circle [radius=.25pt];
\draw (PiS) ++(130:0.4cm) arc (130:180:0.4cm);
 \path [name intersections={of = annulus1 and circle_small_help}];
 \coordinate (P1) at (intersection-1);
 \getx{P1}
 \draw (P1) -- (\mydim,0) node[right,below,xshift=1pt] {$h_i(\hat{\hat{\bS}})$};
\fill (\mydim,0) circle [radius=.25pt];
\draw (PiS) ++(125:0.2cm) arc (125:180:0.2cm);
 \draw [name intersections={of=circle1 and circle_small_help, by=Shat}] (Shat) node[right,xshift=4pt,yshift=-4pt] {$\hat{\hat{\bS}}$};
 \fill (Shat) circle [radius=.25pt];
\draw (0,0) -- (S);
\draw (0,0) -- (Shat);
\draw (0,0) +(30:.3cm) arc(30:41.5:.3cm) node[right,xshift=2pt,yshift=-2pt] {$\beta$};
\draw (0,0) +(41.5:.4cm) arc(41.5:53:.4cm) node[right,xshift=1pt,yshift=1pt] {$\gamma_1$};
\getxxx{PiS}
 \draw (PiS) -- (\mydimmm, 0) node[right,below,xshift=10pt,yshift=-5pt] {$P_i(\bS)'$};
\draw[<->] (\mydim,0.1) -- (\mydimm,0.1) node[below,xshift=20pt,yshift=15pt] {$a$};
\draw[<->] (\mydim,0.32) -- (\mydimmm,0.32) node[below,fill=white!30,xshift=-5pt,yshift=-4pt] {\small{$\norm{\hat{\hat{\bS}}-P_i(\hat{\hat{\bS}})}$}};
\end{tikzpicture}	
\captionof{figure}{Deriving a bound on $\angle{(\hat{\hat{\bS}},\bS)}$ in \eqref{eq:tri_ineq} for the maximum covering angle $\theta$.}\label{graph:theta_part2}
\end{center}
\vspace{\baselineskip}

\vspace{\baselineskip}
For the second part in \eqref{eq:tri_ineq} we consider the setting shown in 
Figure~\ref{graph:theta_part2} and state that the angle~$\beta$ can be calculated as

\begin{align}
 \beta = 2\arcsin\left(\frac{\norm{\hat{\hat{\bS}}-P_i(\hat{\hat{\bS}})}}{2}\right)\label{eq:beta},
\end{align}
and $\gamma_1$ is given by

\begin{align}\label{eq:gamma_big}
 \gamma_1 &= 2\arcsin\left(\frac{\norm{\hat{\hat{\bS}}-P_i(\hat{\hat{\bS}})}+a}{2}\right)- \beta \leq 2\arcsin\left(\frac{\norm{\hat{\hat{\bS}}-P_i(\hat{\hat{\bS}})}+ r_{\Lambda}}{2}\right)- \beta\nonumber\\
          &\leq 2\arcsin\left(\frac{\norm{\hat{\hat{\bS}}-P_i(\hat{\hat{\bS}})}+ r_{\Lambda}}{2}\right) - 2\arcsin\left(\frac{\norm{\hat{\hat{\bS}}-P_i(\hat{\hat{\bS}})}}{2}\right),
\end{align}

where $\norm{\hat{\hat{\bS}}-P_i(\hat{\hat{\bS}})} = 2 \sin{\frac{\arcsin{\hat{x}_n} - \alpha_i}{2}}$ and 
therefore the expression is only a function of $\hat{\bS}$ and index $i$.

Depending on the relation of the latitudes of $\bS$ and $\hat{\hat{\bS}}$ (i.e. $\bS_n \geq \hat{\hat{\bS}}_n$ 
or $\bS_n < \hat{\hat{\bS}}_n$)  a second case has to be taken into account: When $\bS$ is closer to the corresponding annulus 
point $P_i(\bS)$ than $\hat{\hat{\bS}}$ this leads to

\begin{align}\label{eq:gamma_small}
 \gamma_2 &= \beta - 2\arcsin\left(\frac{\norm{\hat{\hat{\bS}}-P_i(\hat{\hat{\bS}})}-a}{2}\right) \leq \beta - 2\arcsin\left(\frac{\norm{\hat{\hat{\bS}}-P_i(\hat{\hat{\bS}})} - r_{\Lambda}}{2}\right)\nonumber\\
          &\leq 2\arcsin\left(\frac{\norm{\hat{\hat{\bS}}-P_i(\hat{\hat{\bS}})}}{2}\right) - 2\arcsin\left(\frac{\norm{\hat{\hat{\bS}}-P_i(\hat{\hat{\bS}})}- r_{\Lambda}}{2}\right).
\end{align}

We observe that \eqref{eq:gamma_small} is always smaller than \eqref{eq:gamma_big} because of the special structure of 
the expressions and the fact that $\frac{\mathrm{d}}{\mathrm{d}x}\arcsin(x) = \frac{1}{\sqrt{1-x^2}}$ is increasing with
larger values of $x$. We therefore set~$\gamma=\max(\gamma_1,\gamma_2)=\gamma_1$ and have derived our desired result.

\noindent \textit{Remark 7: }The bound on the covering angle holds also for the case that a point is quantized in the outside 
of $\cB^{n-1}$ and has to be projected back to the sphere (see remark 3). The reason behind that is that projecting a point in
the outside of $\cB^{n-1}$ back to the sphere makes it closer to the original point. This can easily be shown using the law of 
cosine.

\bibliographystyle{IEEEtran}
\bibliography{main}

\end{document}